\documentclass[aps,pre,twocolumn,groupedaddress]{revtex4-1}

\usepackage[utf8]{inputenc}
\usepackage{amsmath,amsfonts,amssymb}
\usepackage{graphicx}

\usepackage{hyperref}
\usepackage[usenames,dvipsnames]{xcolor}
\hypersetup{colorlinks=true, linkcolor=BrickRed, urlcolor=blue!50!black, citecolor=blue!50!black}

\newcommand{\img}{\mathsf{i}}
\newcommand\diff{\mathrm{d}}
\renewcommand{\vec}[1]{\mathbf{#1}}
\renewcommand{\phi}[0]{\varphi}

\usepackage[normalem]{ulem}

\begin{document}

\title{Time-dependent perpendicular fluctuations in the driven lattice Lorentz gas}



\author{Sebastian Leitmann}
\affiliation{Institut f\"ur Theoretische Physik, Universit\"at Innsbruck, Technikerstra{\ss}e~21A, A-6020 Innsbruck, Austria}
\author{Thomas Schwab}
\affiliation{Institut f\"ur Theoretische Physik, Universit\"at Innsbruck, Technikerstra{\ss}e~21A, A-6020 Innsbruck, Austria}
\author{Thomas Franosch}
\affiliation{Institut f\"ur Theoretische Physik, Universit\"at Innsbruck, Technikerstra{\ss}e~21A, A-6020 Innsbruck, Austria}
\email[]{thomas.franosch@uibk.ac.at}

\date{\today}

\begin{abstract}
We present results for the fluctuations of the displacement of a tracer particle on a planar lattice pulled by a step
force in the presence of impenetrable, immobile obstacles. The fluctuations perpendicular to the applied force are
evaluated exactly in first order of the obstacle density for arbitrarily strong pulling and all times. The complex
time-dependent behavior is analyzed in terms of the diffusion coefficient, local exponent, and the non-Skellam
parameter, which quantifies deviations from the dynamics on the lattice in the absence of obstacles. The non-Skellam parameter
along the force is analyzed in terms of an asymptotic model and reveals a power-law growth for intermediate times.
\end{abstract}


\maketitle



In experiments of active microrheology, a tracer particle is pulled through an environment by optical or magnetic
tweezers~\cite{Squires:LM_24:2008,Wilson:PCCP_13:2011,Puertas:JPCM_26:2014}. The goal is to infer material properties
of the environment in the nonlinear regime not accessible by merely monitoring the thermally agitated motion of a particle
as in passive microrheology~\cite{Mason:PRL_74:1995}. In the presence of pulling, these systems are
driven strongly out of equilibrium, and new phenomena such as
force-thinning~\cite{Habdas:EPL_67:2004,Carpen:JRHEO_49:2005,Sriram:PoF_22:2010}, (transient) superdiffusive behavior,
and enhanced diffusion~\cite{Winter:PRL_108:2012,Winter:JCP_138:2013,Horbach:EPJST_226:2017} emerge.

The nonlinear regime of such systems can be investigated theoretically by considering generic models with a strong
repulsive interaction between the tracer and the particles comprising the environment. For the motion of the tracer, one
conventionally considers Brownian motion in continuum or random walks on lattices. For these cases, the environment can
consist of dilute and immobile obstacles up to crowded environments with a certain underlying dynamics.  For lattice
models, progress and even exact results have been
presented~\cite{Jack:PRE_78:2008,Benichou:PRL_111:2013,Illien:PRL_111:2013,Leitmann:PRL_111:2013,Basu:JPAMT_47:2014,Illien:PRL_113:2014,
Benichou:PRL_113:2014,Illien:JSMTE:2015,Baiesi:PRE_92:2015,Benichou:PRE_93:2016,Leitmann:PRL_118:2017}, whereas in
continuum, continuous-time random walks~\cite{Schroer:PRL_110:2013,Schroer:JCP_138:2013,Burioni:CTP_62:2014}, Langevin
equations~\cite{Demery:NJP_16:2014,Demery:PRE_91:2015}, kinetic theory~\cite{Wang:PRE_93:2016}, and the framework of
mode-coupling theory of the glass transition~\cite{Gazuz:PRL_102:2009,Gnann:SM_7:2011,Gnann:PRE_86:2012,
Harrer:JPCM_24:2012,Gazuz:PRE_87:2013,Wang:PRE_89:2014,Gruber:PRE_94:2016} have been successfully employed.  For active
microrheology in suspensions of hard spheres performing Brownian motion~\cite{Squires:PoF_17:2005,Khair:JFM_557:2006,
Zia:JFM_658:2010,Swan:PoF_25:2013,Hoh:JFM_795:2016}, exact results have been obtained in first order of the density for
the stationary mobility~\cite{Squires:PoF_17:2005} and the stationary diffusion coefficient parallel and perpendicular
to the field~\cite{Zia:JFM_658:2010}.

Here, we employ a lattice model for a tracer in the presence of quenched disorder realized by immobile and impenetrable
obstacles. At time zero, we switch on a constant step force pulling the tracer and monitor the time-dependent dynamics
and the approach to the stationary state. For this model, it is possible to solve for the complete time-dependent
dynamics in first order of the obstacle density and arbitrarily strong driving. Previously, we have discussed the
time-dependent velocity and the growth of the fluctuations along the force given by the variance of the displacement of
the tracer particle~\cite{Leitmann:PRL_111:2013, Leitmann:PRL_118:2017}. Here, we extend and elaborate the solution for
the case of the fluctuations perpendicular to the applied force on the tracer encoded in the respective mean-square
displacement. We characterize the time-dependent dynamics in terms of the diffusion coefficient, the local exponent
encoding sub- and superdiffusive behavior, and the non-Skellam parameter, which encodes deviations from the free motion
of the tracer on the lattice similar to the non-Gaussian parameter for Brownian motion in continuum.

The main results of this work can be summarized in the following way: In equilibrium, the time-dependent diffusion
coefficient is a monotonically decreasing function. This is no longer the case in the presence of a field where the
perpendicular diffusion coefficient shows both a decrease as well as an increase over time. The time-dependent behavior
of the perpendicular diffusion coefficient is observed in the local exponent where transiently subdiffusive and
superdiffusive regimes characterize the approach to the stationary state.  These subdiffusive and superdiffusive regimes
become visible in the non-Skellam parameter as positive and negative contributions. In the stationary state, the
diffusion coefficient perpendicular to the applied force is characterized by density-induced nonanalytic contributions
for small driving. The diffusion coefficient  increases monotonically with increasing force and is bounded from above in
the limit of strong driving.

This work is organized as follows. In section~\ref{sec:the_model} the driven lattice Lorentz gas is defined, and the
notation used throughout this work is introduced. The general solution strategy relying on a scattering theory to
account for repeated encounters of the tracer with the same obstacle is elaborated in
section~\ref{sec:solution_strategy}, and the formal solution is presented in section~\ref{sec:solution}. Readers whose
primary concern is about results rather than the theoretical techniques may skip these sections upon first reading and
jump directly to section~\ref{sec:moments_perpendicular}, where the main results are presented. The discussion
is followed by a summary and conclusion in section~\ref{sec:summary_conclusion}.

\section{The model} \label{sec:the_model}

We consider a tracer particle performing a random walk on a square lattice of lattice spacing $a$ that we set to unity:
$\Lambda = \{\vec{r} = (x, y) \in \mathbb{Z}\times \mathbb{Z} : x, y \in [-L/2, L/2[\}$, linear size $L \in 2\mathbb{N}$ with periodic
boundary conditions and $N = L^2$ sites. The random walker performs successive jumps to its nearest-neighbors, $\mathcal{N} =
\{\pm \vec{e}_x, \pm \vec{e}_y\}$. The lattice consists of free sites, accessible to the tracer as well as sites with
randomly placed immobile hard obstacles of density $n$ (fraction of excluded sites). If the tracer attempts to jump onto an obstacles site, it
remains at its initial position before the jump.  
The waiting time of the tracer at every site is Poisson-distributed with mean waiting time $\tau$.


Statistical information about the random walk is encoded in the site-occupation probability density. Since the time evolution
is described by a linear master equation, it is convenient to adopt a bra-ket notation. We define an abstract ket $|p(t)\rangle$ encoding the 
site-occupation probability density, which can then be expanded in the complete and orthonormal basis of all position kets
$\{|\vec{r}\rangle : \vec{r} \in \Lambda\}$: 
\begin{align}
|p(t)\rangle = \sum_{\vec{r}\in\Lambda} |\vec{r}\rangle \langle \vec{r} | p(t)\rangle. 
\end{align}
Hence, the probability to find the random walker at time $t$ at site $\vec{r}$ is given by the overlap $\langle \vec{r}|p(t)\rangle$.
The evolution in time of the density $|p(t)\rangle$ is determined by the master equation $\partial_t |p(t)\rangle =
\hat{H}|p(t)\rangle$ with ``Hamiltonian'' $\hat{H}$.
In the position basis it obtains the following form:
\begin{align}
\partial_t \langle \vec{r} | p(t) \rangle = 
  \sum_{\vec{r}' \in \Lambda} \langle\vec{r}|\hat{H}|\vec{r}'\rangle\langle\vec{r}'|p(t)\rangle, 
\end{align}
where the matrix elements $\langle \vec{r}|\hat{H}|\vec{r}'\rangle$ encode the transition rates from
site $\vec{r}'$ to $\vec{r}$. 

First we consider the reference case in which there are no impurities on the lattice and every site is accessible to the
tracer. Driving is introduced via a force that pulls the tracer along the $x$-direction of the
lattice. We measure the strength of the force in the dimensionless number $F = \text{force}\cdot(\text{lattice spacing})/k_\text{B} T$. The force
introduces a bias in the corresponding nearest-neighbor transition probabilities $W(\vec{d}\in\mathcal{N})$, and local detailed balance
$W(\vec{e}_x)/W(-\vec{e}_x)=\exp(F)$ and $W(\vec{e}_y)/W(-\vec{e}_y)=1$ along both lattice directions suggests 
\begin{align}
W(\pm \vec{e}_x) = \frac{e^{\pm F/2}}{e^{F/2} + e^{-F/2} + 2}, 
\end{align}
\begin{align}
W(\pm \vec{e}_y) = \frac{1}{e^{F/2} + e^{-F/2} + 2}.
\end{align}
We consider non-normalized rates $(\Gamma/\tau) W(\vec{d}\in\mathcal{N})$ with dimensionless rate $\Gamma = [\cosh(F/2)
+ 1]/2$, and the mean waiting time $\tau$ sets the time scale.
This reference case then defines the unperturbed Hamiltonian 
\begin{align}
\hat{H}_0 = \frac{\Gamma}{\tau}\sum_{\vec{r} \in \Lambda}\Bigl[-|\vec{r}\rangle\langle\vec{r}| +
\sum_{\vec{d}\in\mathcal{N}}W(\vec{d})|\vec{r}\rangle\langle\vec{r}-\vec{d}|\Bigr]. 
\end{align}

In the presence of hard obstacles, transitions to and from impurities are prohibited, which can be formally accounted
for by writing the Hamiltonian as $\hat{H} = \hat{H}_0 + \hat{V}$, such that the ``potential'' $\hat{V}$ cancels the
forbidden transitions.  In particular, for a single impurity at site $\vec{s}_1$, we obtain $\hat{V} =
\hat{v}(\vec{s}_1) \equiv \hat{v}_1$ and the only non-vanishing matrix elements affect the obstacle site $\vec{s}_1$ and its
nearest-neighbors $\vec{s}_1 - \vec{d}$, with $\vec{d} \in\mathcal{N}$ leading to $\langle \vec{s}_1 | \hat{v}_1 |
\vec{s}_1\rangle = (\Gamma/\tau)$, $\langle \vec{s}_1 | \hat{v}_1 | \vec{s}_1 - \vec{d}\rangle
= -(\Gamma/\tau) W(\vec{d})$, $\langle \vec{s}_1 - \vec{d} | \hat{v}_1 | \vec{s}_1\rangle = -(\Gamma/\tau) W(-\vec{d})$, and $\langle
\vec{s}_1 - \vec{d} | \hat{v}_1 | \vec{s}_1 - \vec{d}\rangle = (\Gamma/\tau) W(\vec{d})$.  The complete potential $\hat{V}$ for
$N_I$ impurities is then obtained in first order of the density $n = N_I/N$ by summing over all single-obstacles
potentials: $\hat{V} = \sum_{i = 1}^{N_I} \hat{v}_i$.

The force on the tracer is switched on a time $t = 0$, and we use the thermal equilibrium state $|p(t=0)\rangle =
|p_\text{eq}\rangle = 1/N$ in the absence of driving $F = 0$ as the initial condition. Formally, the tracer is allowed to
start also at an impurity site, however these contributions can be exactly corrected in first order of the density at the
end of the calculation.

\section{Solution strategy} \label{sec:solution_strategy}

We first solve for the dynamics of a particle in the absence of obstacles and express
the time-dependent site-occupation probability density in terms of the time-evolution operator $\hat{U}_0(t)$ via 
$|p(t)\rangle = \hat{U}_0(t)|p_\text{eq}\rangle$. The time-evolution operator for the free dynamics fulfills the
differential equation $\partial_t \hat{U}_0(t) = \hat{H}_0\hat{U}_0(t)$ with initial condition $\hat{U}_0(0) = \openone$ and is
given by $\hat{U}_0(t) = \exp(\hat{H}_0 t)$.
Since the free Hamiltonian $\hat{H}_0$ is translationally invariant, it is diagonal in the plane-wave basis  defined by 
\begin{align}
|\vec{k}\rangle = \frac{1}{\sqrt{N}}\sum_{\vec{r}\in\Lambda}\exp(\img\vec{k}\cdot\vec{r})|\vec{r}\rangle ,
\end{align}
with wave vector $\vec{k} = (k_x, k_y) \in \Lambda^* = \{(2\pi x/L, 2\pi y/L) : (x, y) \in \Lambda\}$, and
scalar product $\vec{k}\cdot\vec{r} = k_x x + k_y y$. 
Then, the invariance under translation implies
$\langle \vec{k} | \hat{H}_0 | \vec{k}'\rangle = \epsilon(\vec{k})\delta(\vec{k},\vec{k}')$ with Kronecker-Delta
$\delta(\vec{k},\vec{k}')$ and eigenvalue $\epsilon(\vec{k})$ of the free Hamiltonian:
\begin{align}
\epsilon(\vec{k}) = -\frac{\Gamma}{\tau}\sum_{\vec{d}\in\mathcal{N}} \bigl[\bigl(1-\cos(\vec{k}\cdot\vec{d})\bigr) +
\img\sin(\vec{k}\cdot\vec{d})\bigr]W(\vec{d}) .
\end{align}
Hence, the time-evolution operator for the free dynamics, $\hat{U}_0(t) = \exp(\hat{H}_0 t)$, is also diagonal in the
plane-wave basis with $\langle \vec{k}|\hat{U}_0(t)|\vec{k}'\rangle = \exp[\epsilon(\vec{k})t]\delta(\vec{k},\vec{k}')$.

All moments of the time-dependent displacement in the absence of obstacles are contained in the self-intermediate scattering function
$F_0(\vec{k},t) = \langle e^{-\img\vec{k}\cdot\Delta\vec{r}(t)}\rangle_0$, which is defined in terms of the
time-evolution operator $\hat{U}_0(t)$ via 
\begin{align}
\begin{split}
F_0(\vec{k},t) 
               &= \sum_{\vec{r},\vec{r}'\in\Lambda} e^{-\img\vec{k}\cdot(\vec{r}-\vec{r}')}
                  \langle\vec{r}|\hat{U}_0(t)|\vec{r}'\rangle\langle\vec{r}'|p_\text{eq}\rangle,  
\end{split}
\end{align}
with initial distribution $\langle \vec{r}'|p_\text{eq}\rangle = 1/N$.
It is directly connected to the eigenvalue of the time-evolution operator in the plane-wave basis with 
\begin{align}
F_0(\vec{k},t) = \langle\vec{k}|\hat{U}_0(t)|\vec{k}\rangle = \exp[\epsilon(\vec{k}) t]. 
\end{align}

In the presence of obstacles, we express the dynamics in terms of the scattering formalism borrowed from quantum
mechanics~\cite{Ballentine:WS:2003}. We define the propagator $\hat{G}$ by the Laplace transform of the respective time-evolution operator
$\hat{U}(t)$ for a configuration $\hat{V} = \sum_{i=1}^{N_I}\hat{v}_i$ of obstacles:
\begin{align} \label{eq:definition_propagator}
\hat{G}(s) = \int_0^\infty \diff t\ e^{-s t} \hat{U}(t) = (s - \hat{H})^{-1} .
\end{align}
In particular, the free propagator $\hat{G}_0 = (s - \hat{H}_0)^{-1}$ is diagonal in the plane-wave basis with the eigenvalue  
\begin{align} \label{eq:free_propagator_plane_wave}
G_0(\vec{k}) = \langle \vec{k} | \hat{G}_0 | \vec{k} \rangle = \frac{1}{s - \epsilon({\vec{k}})}\ .
\end{align}
The dependence on the Laplace frequency $s$ will be suppressed throughout.

The scattering operator $\hat{T} = \hat{V} + \hat{V}\hat{G}_0\hat{T}$ accounts for all possible collision events of the
tracer with the obstacle disorder and connects both propagators via the relation
\begin{align}
\hat{G} = \hat{G}_0 + \hat{G}_0 \hat{T} \hat{G}_0.
\end{align}
Inserting the obstacle configuration $\hat{V} = \sum_{i=1}^{N_I}\hat{v}_i$ into the scattering operator
expansion $\hat{T} = \hat{V} + \hat{V}\hat{G}_0\hat{V} + \hat{V}\hat{G}_0\hat{V}\hat{G}_0\hat{V} + \dotsb$, 
one arrives at
\begin{align}
\begin{split}
\hat{T} = \sum_{i=1}^{N_I} \hat{v}_i + \sum_{j,k=1}^{N_I}\hat{v}_j \hat{G}_0 \hat{v}_k
+ \sum_{l,m,n=1}^{N_I} \hat{v}_l\hat{G}_0\hat{v}_m\hat{G}_0\hat{v}_n + \dotsb .
\end{split}
\end{align}
The possible scattering events can be arranged in terms of repeated collisions with the same obstacle $\hat{v}_i
\hat{G}_0 \hat{v}_i$ and distinct collisions $\hat{v}_i \hat{G}_0 \hat{v}_j$ with two different obstacles $i \neq j$. This
classification is conveniently expressed in terms of the scattering operator of a single obstacle,
\begin{align}
\hat{t}_i = \hat{v}_i + \hat{v}_i\hat{G}_0\hat{v}_i + \hat{v}_i \hat{G}_0 \hat{v}_i \hat{G}_0 \hat{v}_i + \dotsb ,
\end{align}
which accounts for all possible repeated collisions of the tracer with the same obstacle $\hat{v}_i$.
The scattering operator $\hat{T}$ can then be written in terms of these scattering operators $\hat{t}_i$, leading
to the multiple scattering expansion
\begin{align} \label{eq:scattering_expansion_t}
\hat{T} = \sum_{i=1}^{N_I} \hat{t}_i + \sum_{\substack{j,k=1\\ j\neq k}}^{N_I}\hat{t}_j \hat{G}_0 \hat{t}_k
+ \sum_{\substack{l,m,n=1 \\ l\neq m, m\neq n}}^{N_I} \hat{t}_l\hat{G}_0\hat{t}_m\hat{G}_0\hat{t}_n + \dotsb .
\end{align}

Since here we are not interested in a particular configuration of obstacles, we take an average over the disorder
realizations $[.]_\text{av}$, which also restores translational invariance. Then we evaluate the scattering operator in
the plane-wave basis, $\langle \vec{k} | [\hat{T}]_\text{av} | \vec{k} \rangle$, where only diagonal elements are
non-vanishing. The first sum in the scattering expansion [Eq.~(\ref{eq:scattering_expansion_t})] is then identified as
contributions in first order of the density $n$ and is called the independent-scatterer
approximation~\cite{Rossum:RMP_71:1999}. The remaining contributions are of order $n^2$ or higher and describe
correlated scattering events between different obstacles.

An exact expression for the disorder-averaged propagator in the plane-wave basis, $[G]_\text{av}(\vec{k}) :=
\langle \vec{k} |[\hat{G}]_\text{av}|\vec{k}\rangle$, in first order of the density of the obstacles $n$ is then obtained
as 
\begin{align}
[G]_\text{av}(\vec{k}) =  G_0(\vec{k}) + nN t(\vec{k}) G_0(\vec{k})^2 + \mathcal{O}(n^2) , 
\end{align}
with the forward-scattering amplitude $t(\vec{k}) = \langle \vec{k} | \hat{t} | \vec{k} \rangle$ of a single obstacle.
Since the forward-scattering amplitude is itself of order $\mathcal{O}(1/N)$, the disorder-averaged propagator converges
in the limit of large lattices $L \to \infty$.  
The disorder-averaged propagator is connected to the self-energy $\Sigma(\vec{k})$ in terms of the Dyson equation
\begin{align}
[G]_\text{av}(\vec{k}) = \frac{1}{G_0(\vec{k})^{-1} - \Sigma(\vec{k})}.
\end{align}
Thus, the contributions to the self-energy in first order of the density are encoded in the scattering $t$-matrix
via $\Sigma(\vec{k}) = nNt(\vec{k}) + \mathcal{O}(n^2)$. Similar to the obstacle-free case, a temporal Laplace
transform of the intermediate scattering function,
\begin{align}
\begin{split}
F(\vec{k},t) &= \langle e^{-\img\vec{k}\cdot\Delta\vec{r}(t)}\rangle \\
            &= \sum_{\vec{r},\vec{r}'\in\Lambda} e^{-\img\vec{k}\cdot(\vec{r}-\vec{r}')}
                  \langle\vec{r}|[\hat{U}(t)]_\text{av}|\vec{r}'\rangle\langle\vec{r}'|p_\text{eq}\rangle , 
\end{split}
\end{align}
leads to the disorder-averaged propagator $[G]_\text{av}(\vec{k})$. The moments of displacement can then be
obtained in the frequency domain as certain derivatives with respect to the wave vector $\vec{k}$. In our calculation, the tracer
particle is allowed to start at an obstacle site where it remains forever. We trivially correct for this behavior by multiplying the
disorder-averaged propagator $[G]_\text{av}(\vec{k})$ with $1/(1-n) = 1 + n + \mathcal{O}(n^2)$ and keeping only
contributions in first order of the density. Hence, we obtain
\begin{align} \label{eq:disorder_averaged_prop_corrected}
[G]_\text{av}^\text{c}(\vec{k}) =  G_0(\vec{k}) + n[G_0(\vec{k}) + N t(\vec{k}) G_0(\vec{k})^2] + \mathcal{O}(n^2) . 
\end{align}

The remaining task is the calculation of the forward-scattering amplitude $t(\vec{k})$. The scattering operator
for a single obstacle $\hat{v}$ fulfills the relation 
\begin{align} \label{eq:identity_single_scattering_t}
\hat{t} = \hat{v} + \hat{v}\hat{G}_0\hat{t} = \hat{v} + \hat{t}\hat{G}_0\hat{v}. 
\end{align}
Due to nearest-neighbor hopping, the only non-vanishing matrix elements $\langle \vec{r} | \hat{t} | \vec{r}'
\rangle$ in the real-space basis $\vec{r}$ correspond to the distinguished subspace consisting of the location of the obstacle, which we put at
the origin $\vec{0}$, and its
neighboring sites $\mathcal{N}$. Thus, the operator identity
[Eq.~(\ref{eq:identity_single_scattering_t})] can be read as a $5\times 5$ matrix inversion problem, and we can
restrict our calculations to the distinguished subspace spanned by the impurity site and its neighbors. We introduce the basis
$\{\vec{e}_{-2}, \vec{e}_{-1}, \vec{e}_0, \vec{e}_1, \vec{e}_2\}$, where we identify $\vec{e}_{\pm 2} := \pm \vec{e}_{y}$, 
$\vec{e}_{\pm 1} := \pm \vec{e}_{x}$, and $\vec{e}_0 := \vec{0}$ such that the single-obstacle potential $\hat{v}$ takes
the form 
\begin{align} \label{eq:obs_potential_site}
v = \frac{1}{4\tau}
\begin{pmatrix}
1 & 0 & -1 & 0 & 0 \\
0 & e^{F/2} & -e^{-F/2} & 0 & 0 \\
-1 & -e^{F/2} & 4\Gamma & -e^{-F/2} & -1 \\
0 & 0 & -e^{F/2} & e^{-F/2} & 0 \\
0 & 0 & -1 & 0 & 1 
\end{pmatrix} .
\end{align}
Note that the sum in each column evaluates to zero, which reflects the conservation of probability.
For the matrix form of the propagator $\hat{G}_0$ in real space $\vec{r}$, we start with the exact solution 
of the time-evolution operator in the limit of large lattices $L \to \infty$~\cite{Montroll:JSP_9:1973,Haus:PR_150:1987}: 
\begin{align} \label{eq:sol_time_evolution_op}
\langle \vec{r}|\hat{U}_0(t)|\vec{r}'\rangle = \exp(F \Delta x/2) \langle \vec{r} | \hat{u}_0(t) |\vec{r}'\rangle,
\end{align}
where we introduced the symmetric propagator 
\begin{align} \label{eq:symmetric_propagator}
\langle \vec{r}|\hat{u}_0(t)|\vec{r}'\rangle = e^{-\Gamma t/\tau} I_{\Delta x}(t/2\tau) I_{\Delta y}(t/2\tau),
\end{align}
with modified Bessel function $I_m(\cdot)$ of integer order $m$. Note that the symmetric propagator still depends on
the force via the dimensionless rate $\Gamma$.  
We define the
Laplace transform of the symmetric part of the time-evolution operator $\hat{g}(s) = \int_0^\infty \diff t\ e^{-s t}
\hat{u}(t)$, and we obtain the matrix form of the free propagator in the basis
$\{\vec{e}_{-2},\vec{e}_{-1},\vec{e}_0,\vec{e}_1,\vec{e}_2\}$ as
\begin{align}
G_0 = \begin{pmatrix}
g_{00} & e^{F/2} g_{11} & g_{10} & e^{-F/2} g_{11} & g_{20} \\
e^{-F/2} g_{11} & g_{00} & e^{-F/2} g_{10} & e^{-F} g_{20} & e^{-F/2} g_{11} \\
g_{10} & e^{F/2} g_{10} & g_{00} & e^{-F/2} g_{10} & g_{10} \\
 e^{F/2} g_{11} & e^{F} g_{20} & e^{F/2} g_{10} & g_{00} & e^{F/2} g_{11} \\
g_{20} & e^{F/2} g_{11} & g_{10} & e^{-F/2} g_{11} & g_{00} 
\end{pmatrix} ,
\end{align}
where we abbreviated the propagators for coming back to the same site, $g_{00} := \langle \vec{e}_0 | \hat{g} |
\vec{e}_0 \rangle$, arriving at a neighboring site, $g_{10} := \langle \vec{e}_1 | \hat{g} | \vec{e}_0 \rangle$, 
arriving at the next-neighboring site along a direction, $g_{20} := \langle \vec{e}_2 | \hat{g} | \vec{e}_{-2} \rangle$,
and 
along a diagonal, $g_{11} := \langle \vec{e}_1 | \hat{g} | \vec{e}_{-2} \rangle$.

\section{Solution} \label{sec:solution}

To illustrate the solution strategy, we reconsider the case of no driving, $F=0$, where a solution was achieved much
earlier in a different way~\cite{Nieuwenhuizen:PRL_57:1986,Nieuwenhuizen:JPAMG_20:1987,Ernst:JPAMG_20:1987}. For the calculation of the
scattering $t$-matrix, we perform a change of basis adapted to the symmetry of the problem in
the case of no driving:
\begin{align}
\begin{split} \label{eq:new_basis_symmetry}
\sqrt{2}|p_y\rangle &= |\vec{e}_{-2}\rangle - |\vec{e}_2\rangle, \\
\sqrt{2}|p_x\rangle &= |\vec{e}_{-1}\rangle - |\vec{e}_1\rangle, \\
2|d_{xy}\rangle &= |\vec{e}_{-2}\rangle - |\vec{e}_{-1}\rangle - |\vec{e}_1\rangle + |\vec{e}_2\rangle, \\
2\sqrt{5}|s\rangle &= |\vec{e}_{-2}\rangle + |\vec{e}_{-1}\rangle - 4|\vec{e}_0\rangle  + |\vec{e}_1\rangle +
|\vec{e}_2\rangle, \\
\sqrt{5} |n\rangle &= |\vec{e}_{-2}\rangle + |\vec{e}_{-1}\rangle + |\vec{e}_0\rangle  + |\vec{e}_1\rangle +
|\vec{e}_2\rangle. 
\end{split}
\end{align}
The first two vectors are reminiscent of dipoles, the third one is of quadrupolar type, whereas the last two are invariant under
rotations. One observes, that the last mode $|n\rangle$ is connected to the conservation of probability via $\langle
n|\hat{v} = 0$ and acts as a neutral mode. The respective transformation 
is encoded in the orthogonal matrix
\begin{align} \label{eq:orth_trans_matrix}
M = \begin{pmatrix}
\frac{1}{\sqrt{2}} & 0 & \frac{1}{2} & \frac{1}{2\sqrt{5}} & \frac{1}{\sqrt{5}} \\
0 & \frac{1}{\sqrt{2}} & -\frac{1}{2} & \frac{1}{2\sqrt{5}} & \frac{1}{\sqrt{5}} \\
0 & 0 & 0 & -\frac{2}{\sqrt{5}} & \frac{1}{\sqrt{5}} \\
0 & -\frac{1}{\sqrt{2}} & -\frac{1}{2} & \frac{1}{2\sqrt{5}} & \frac{1}{\sqrt{5}} \\
-\frac{1}{\sqrt{2}} & 0 & \frac{1}{2} & \frac{1}{2\sqrt{5}} & \frac{1}{\sqrt{5}}
\end{pmatrix} .
\end{align}
The new basis [Eq.~\eqref{eq:new_basis_symmetry}] and thus the orthogonal transform $M$ can be rationalized in the
framework of group theory with respect to the possible symmetry transformation of the dihedral group $D_4$ (see
Appendix~\ref{Appendix:Group_Theory}).

In this new representation, the matrix for the obstacle potential in the absence of driving becomes diagonal with $v' = M^T v M =
(1/4\tau)\text{diag}(1, 1, 1, 5, 0)$. The vanishing of the last row reflects conservation of probability. 
The free propagator in the new basis, $G_0' = M^T G_0 M$, is diagonal up to the nonvanishing entries $\langle s |
\hat{G}_0 | n \rangle$ and $\langle n | \hat{G}_0|s \rangle$. However, the scattering $t$-matrix, becomes diagonal again with 
\begin{align}
\begin{split}
t' &= (1 - v' G_0')^{-1}v' \\ 
&= \text{diag}\Bigl(\frac{1}{4\tau + (g_{20}-g_{00})}, \frac{1}{4\tau + (g_{20}-g_{00})}, \\
&\phantom{==}\frac{1}{4\tau + (2g_{11}-g_{20}-g_{00})},  \frac{5}{4\tau (g_{00}-g_{10})s},0\Bigr).
\end{split}
\end{align}

To determine the matrix elements of the scattering $t$-matrix in the plane-wave basis, we only have to consider
contributions from the distinguished subspace $\vec{r} \in {\vec{0}}\cup\mathcal{N}$. Thus, we decompose the projection of the
wave vector onto the distinguished subspace $\hat{P}|\vec{k}\rangle = (1/\sqrt{N})\sum_{\nu=-2}^2 e^{\img\vec{k}\cdot\vec{e}_\nu}|\vec{e}_\nu\rangle$ in the new basis 
introduced by the transformation $M$~[Eq.~\eqref{eq:orth_trans_matrix}]:
\begin{align}
\begin{split} \label{eq:wave_vector_symmetry_basis}
\sqrt{N}\hat{P}|\vec{k}\rangle =&-\img\sqrt{2}\sin(k_y)|p_y\rangle -\img\sqrt{2}\sin(k_x)|p_x\rangle\\
                 &+[\cos(k_y)-\cos(k_x)]|d_{xy}\rangle\\
                 &+\frac{1}{\sqrt{5}}[\cos(k_x)+\cos(k_y)-2]|s\rangle\\
                 &+\frac{1}{\sqrt{5}}[2\cos(k_x)+2\cos(k_y)+1]|n\rangle .
\end{split}
\end{align}
Since the scattering $t$-matrix for the equilibrium case is diagonal in this basis, the matrix element
$t(\vec{k}) = \langle \vec{k} |\hat{t}|\vec{k}\rangle$ can be readily calculated:
\begin{align}
\begin{split}
N t(\vec{k}) =&\ \frac{2[\sin^2(k_x) + \sin^2(k_y)]}{4\tau + (g_{20} - g_{00})} 
+ \frac{[\cos(k_x) - \cos(k_y)]^2}{4\tau+(2g_{11} - g_{20} - g_{00})} \\
&\ + \frac{[\cos(k_x) + \cos(k_y) - 2]^2}{4\tau (g_{00} - g_{10})s}, \quad\text{for } F = 0,  \\
\end{split}
\end{align}
which in principle can also be obtained from Ref.~\cite{Nieuwenhuizen:PA_157:1989}.

For finite force, $F \neq 0$, the applicable symmetry transformations reduce considerably. However, it is still
advantageous to use the orthogonal matrix $M$~[Eq.~\eqref{eq:orth_trans_matrix}]. Then, the matrices for the obstacle
potential $v' = M^T v M$ and the propagator $G_0' = M^T G_0 M$ contain the following nonvanishing entries indicated
by~$*$:
\begin{align}
v' =
\begin{pmatrix}
* & 0 & 0 & 0 & 0 \\
0 & * & * & * & * \\
0 & * & * & * & 0 \\
0 & * & * & * & 0 \\
0 & 0 & 0 & 0 & 0 \\
\end{pmatrix},\quad
G_0' = 
\begin{pmatrix}
* & 0 & 0 & 0 & 0 \\
0 & * & * & * & * \\
0 & * & * & * & * \\
0 & * & * & * & * \\
0 & * & * & * & * \\
\end{pmatrix}. 
\end{align}
The vanishing of the last row in $v'$ again reflects the conservation of probability $\langle n |\hat{v} = 0$. Moreover, 
the $|p_y\rangle$ mode decouples from the problem by the residual mirror symmetry.
This leads to 
\begin{align}
(1-v'G_0') =
\begin{pmatrix}
* & 0 & 0 & 0 & 0 \\
0 & * & * & * & * \\
0 & * & * & * & * \\
0 & * & * & * & * \\
0 & 0 & 0 & 0 & 1 \\
\end{pmatrix},
\end{align}
such that the scattering $t$-matrix in the driven case assumes the following form:
\begin{align} \label{eq:schematic_t_matrix_nonequilibrium}
t' = (1-v'G_0')^{-1} v' =
\begin{pmatrix}
* & 0 & 0 & 0 & 0 \\
0 & * & * & * & * \\
0 & * & * & * & * \\
0 & * & * & * & * \\
0 & 0 & 0 & 0 & 0 \\
\end{pmatrix}.
\end{align}
The scattering $t$-matrix inherits the structure of the potential $\hat{v}$ reflecting conservation of probability and
decoupling of the $|p_y\rangle$ mode. The evaluation of the scattering $t$-matrix essentially reduces to solving the
$3\times 3$ matrix problem, which can be efficiently implemented in computer algebra.

For the moments, we need the derivatives of the wave vector in the distinguished subspace $\hat{P}|\vec{k}\rangle$ with respect to $k_x$ and $k_y$. They are given by the
following expressions:
\begin{align}
\frac{\partial}{\partial k_x}\sqrt{N}\hat{P}|\vec{k}\rangle\Bigr|_{\vec{k}=0} &= -\img\sqrt{2}|p_x\rangle, \\
\frac{\partial^2}{\partial k_x^2}\sqrt{N}\hat{P}|\vec{k}\rangle\Bigr|_{\vec{k}=0} &= |d_{xy}\rangle - \frac{1}{\sqrt{5}}|s\rangle -
\frac{2}{\sqrt{5}}|n\rangle, 
\end{align}
and similarly
\begin{align}
\frac{\partial}{\partial k_y}\sqrt{N}\hat{P}|\vec{k}\rangle\Bigr|_{\vec{k}=0} &= -\img\sqrt{2}|p_y\rangle,  \\
\frac{\partial^2}{\partial k_y^2}\sqrt{N}\hat{P}|\vec{k}\rangle\Bigr|_{\vec{k}=0} &= -|d_{xy}\rangle - \frac{1}{\sqrt{5}}|s\rangle -
\frac{2}{\sqrt{5}}|n\rangle.
\end{align}
Thus, for the first order derivatives after $k_x$ and $k_y$ and for large wavelength $\vec{k} = 0$, we obtain
\begin{align}
\frac{\partial N t}{\partial k_x}\Bigr|_{\vec{k}=0} &= \img \sqrt{10}\langle p_x | \hat{t} | n\rangle,
\label{eq:t_matrix_dkx1} \\
\frac{\partial N t}{\partial k_y}\Bigr|_{\vec{k}=0} &= \img \sqrt{10}\langle p_y | \hat{t} | n\rangle = 0,
\end{align}
whereas the second order derivatives are given by
\begin{align}
\frac{\partial^2 N t}{\partial k_x^2}\Bigr|_{\vec{k}=0} &= \sqrt{5} \langle d_{xy} | \hat{t} | n\rangle - \langle s|\hat{t}|n\rangle + 4\langle p_x|\hat{t}|p_x\rangle 
\end{align}
and
\begin{align} \label{eq:t_matrix_dky2}
\frac{\partial^2 N t}{\partial k_y^2}\Bigr|_{\vec{k}=0} &= -\sqrt{5} \langle d_{xy} | \hat{t} | n\rangle -
\langle s|\hat{t}|n\rangle + 4\langle p_y|\hat{t}|p_y\rangle .
\end{align}
Expressions of the scattering $t$-matrix for higher-order derivatives can be readily computed by determining the corresponding derivatives with respect to the wave
vector~[Eq.~\eqref{eq:wave_vector_symmetry_basis}] and identifing the nonvanishing matrix entries from
Eq.~\eqref{eq:schematic_t_matrix_nonequilibrium}. Explicit expressions for the matrix elements in
Eqs.~\eqref{eq:t_matrix_dkx1}~-~\eqref{eq:t_matrix_dky2} in terms of the
propagators $g_{00}$, $g_{10}$, $g_{11}$, $g_{20}$ are given in Appendix~\ref{appendix:explicit_expressions_matrix_elements}.

The remaining task is the calculation of the four propagators $g_{00}$, $g_{11}$, $g_{10}$, and $g_{20}$.  They can be
expressed in terms of complete elliptic integrals $\mathsf{K}[m] := \int_0^{\pi/2} \diff \theta [1-m\sin^2(\theta)]^{-1/2}$ and
$\mathsf{E}[m] := \int_0^{\pi/2} \diff \theta [1-m\sin^2(\theta)]^{1/2}$.  For the equilibrium case, $F=0$, these propagators have been known
for a long time~\cite{Wortis:PRE_132:1963}. For finite force, the essential modification consists of a shift in the frequency
$s\mapsto s + (\Gamma-1)/\tau$ as inferred from Eq.~\eqref{eq:symmetric_propagator}. Starting with $g_{11}$ and
$g_{00}$, we obtain~\cite{Gradshteyn:AP:2007}
\begin{align}
\begin{split}
\frac{\pi}{2\tau}g_{11} = &-\frac{2(\Gamma+s\tau)^2 \mathsf{E}[(\Gamma+s\tau)^{-2}]}{\Gamma+s\tau} + \\
&-\frac{[1 - 2(\Gamma+s\tau)^2]\mathsf{K}[(\Gamma+s\tau)^{-2}]}{\Gamma+s\tau}, 
\end{split}
\end{align}
as well as
\begin{align}
\frac{\pi}{2\tau}g_{00} = \frac{\mathsf{K}[(\Gamma +s\tau)^{-2}]}{(\Gamma+s\tau)}. 
\end{align}
The propagators are not independent and are related
to each other by the identities $\langle \vec{e}_0 | (s - \hat{H}_0)\hat{G}_0 | \vec{e}_0\rangle = 1$ and $\langle
\vec{e}_0 | (s - \hat{H}_0)\hat{G}_0 | \vec{e}_2 \rangle = 0$, which follow from the definition of the
propagator~[Eq.~\eqref{eq:definition_propagator}]. Evaluation of these expressions leads to the relations
\begin{align}
\begin{split}
(\Gamma + s\tau)g_{00} - g_{10} = \tau,  \\
4(\Gamma + s\tau)g_{10} - (2g_{11} + g_{20} + g_{00}) = 0, 
\end{split}
\end{align}
from which the remaining propagators $g_{10}$ and $g_{20}$ follow directly as
\begin{align}
\frac{\pi}{2\tau}g_{10} = -\frac{\pi}{2} + \mathsf{K}[(\Gamma + s\tau)^{-2}]
\end{align}
and
\begin{align}
\begin{split}
\frac{\pi}{2\tau}g_{20} = &-2(\Gamma +s\tau)(\pi - 2\mathsf{E}[(\Gamma+s\tau)^{-2}]) + \\
&+\frac{\mathsf{K}[(\Gamma + s\tau)^{-2}]}{\Gamma +s\tau} .
\end{split}
\end{align}

\section{Moments perpendicular to the field} \label{sec:moments_perpendicular}

\subsection{Mean-square displacement}

With the scattering $t$-matrix for a single obstacle in the plane-wave basis $t(\vec{k})$ and the free propagator $G_0(\vec{k}) = [s -
\epsilon(\vec{k})]^{-1}$~[Eq.~\eqref{eq:free_propagator_plane_wave}], all moments of the displacement in first order of the density are
determined by the disorder-averaged propagator $[G]_\text{av}^\text{c}(\vec{k})$~[Eq.~\eqref{eq:disorder_averaged_prop_corrected}]. 
Then, the moments can be obtained by derivatives with respect to the wave vector $\vec{k}$. Since the expressions are lengthy,
computer algebra is advantageous for their evaluation.

\begin{figure}
\includegraphics{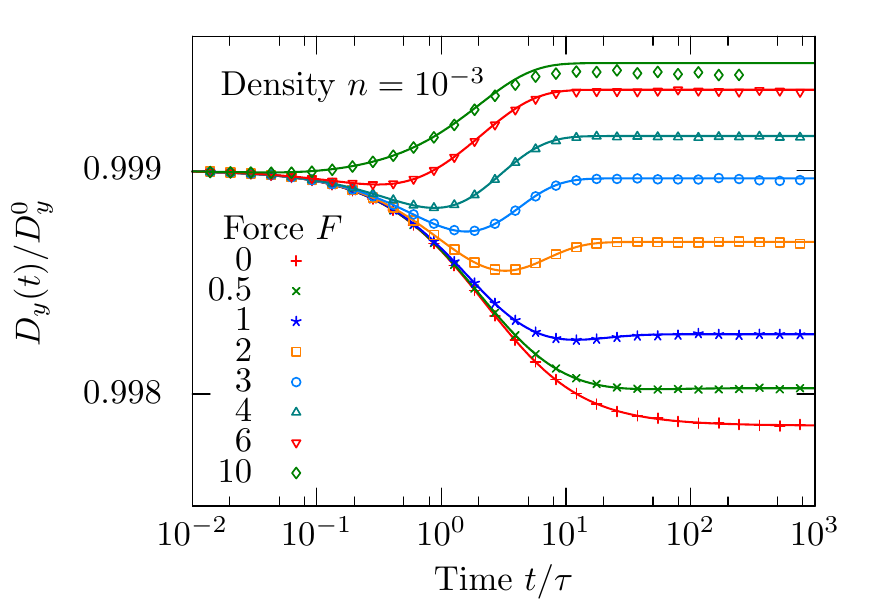}
\caption{\label{fig:diffcoeffy}
Time-dependent diffusion coefficient $D_y(t)$ perpendicular to the force normalized by the diffusion coefficient $D_y^0$
in the absence of obstacles for different dimensionless forces $F$. Solid lines correspond to the analytic solution
and symbols represent stochastic simulations.
}
\end{figure}

For the model considered here, the
velocity and the variance of the tracer displacements along the field have been discussed
recently~\cite{Leitmann:PRL_111:2013,Leitmann:PRL_118:2017}.
Here, we concentrate on the motion perpendicular to the field. Since the random walk of the tracer is symmetric with
respect to the $y$ axis, the displacement $\Delta y(t) = y(t) - y(0)$ along that direction vanishes in the mean,
$\langle \Delta y(t) \rangle = 0$. The first nonvanishing moment of the displacement $\Delta y(t)$ is given by the
mean-square displacement $\langle \Delta y(t)^2\rangle$, which is obtained in the frequency domain via  
\begin{align}
\begin{split}
-\frac{\partial^2}{\partial k_y^2} [G]_\text{av}^\text{c}\biggr|_{\vec{k}=0}
=-\frac{\partial^2 G_0}{\partial k_y^2}\biggr|_{\vec{k}=0}\!\!\! - n \biggl[\frac{\partial^2 G_0}{\partial k_y^2} +
G_0^2 \frac{\partial^2 N t}{\partial k_y^2}\biggr]_{\vec{k}=0},
\end{split}
\end{align}
where $\partial^2 G_0/\partial k_y^2|_{\vec{k}=0} = -2 D_y^0/s^2$ with bare diffusion coefficient $D_y^0 = 1/4\tau$, and $\partial^2 N t/\partial k_y^2|_{\vec{k}=0}$ is given in Eq.~\eqref{eq:t_matrix_dky2}

\begin{figure}
\includegraphics{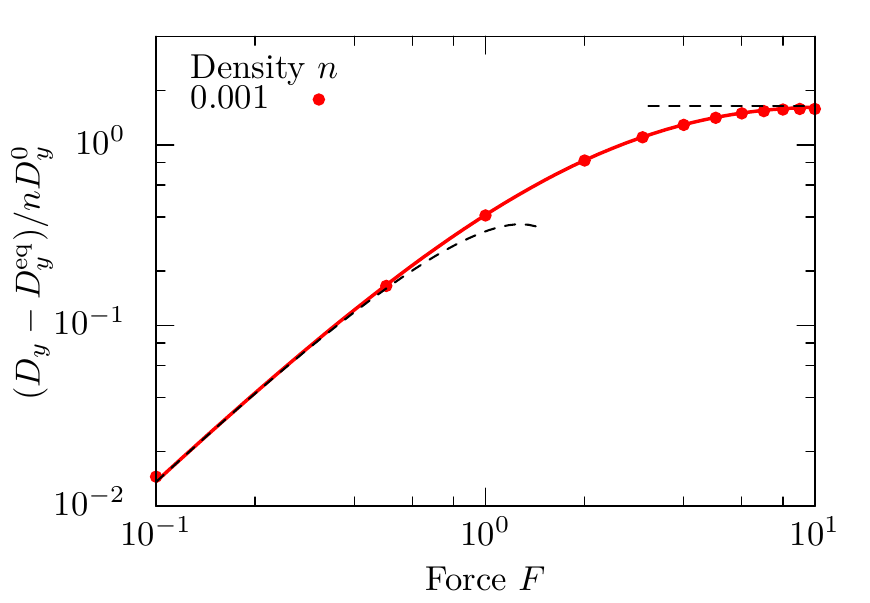}
\caption{\label{fig:stat_diffcoeffy}
Force-induced diffusion coefficient perpendicular to the applied force, $D_y - D_y^\text{eq}$. Solid lines correspond to
the analytic solution and symbols represent stochastic simulations. The dashed lines indicate the limiting behavior for
small~[Eq.~\eqref{eq:diffcoeff_y_small_driving}] and large driving, $D_y(F\to\infty) = D_y^0 (1 - n/2)$.
}
\end{figure}

In the absence of obstacles the mean-square displacement perpendicular to the field reduces to $\langle \Delta
y(t)^2\rangle_0 = 2 D_y^0 t$.
Deviations from the bare case without obstacles can be characterized by the time-dependent diffusion
coefficient perpendicular to the field defined via
\begin{align}
D_y(t) := \frac{1}{2}\frac{\diff}{\diff t} \langle \Delta y(t)^2\rangle .
\end{align}
For short times $t \to 0$, the time-dependent diffusion coefficient is solely determined by the first jump event of the
random walker leading to $D_y(t\to 0) = D_y^0(1 - n)$~[Fig.~\ref{fig:diffcoeffy}]. 
For vanishing force $F = 0$, we recover the analytic solution for the time-dependence of
the perpendicular diffusion coefficient in terms of the Laplace transform
$\hat{D}_y(s, F) = \int_0^\infty \diff t\ e^{-st} D_y(t, F)$:
\begin{align}
\hat{D}_y^\text{eq}(s) = \frac{D_y^0}{s} + n\frac{D_y^0}{s}\biggl[1 - \frac{2}{1 + D_y^0(g_{20} -
g_{00})}\biggr],
\end{align}
with $\hat{D}_y^\text{eq}(s) := \hat{D}_y(s,F=0)$ and the propagators $g_{20}$ and $g_{00}$ in equilibrium ($\Gamma = 1$), which have been calculated
earlier~\cite{Nieuwenhuizen:PRL_57:1986,Nieuwenhuizen:JPAMG_20:1987,Ernst:JPAMG_20:1987}.
The long-time behavior of the time-dependent diffusion coefficient is encoded in the respective small-frequency
behavior $s\to 0$: 
\begin{align} \label{eq:diff_coeff_eq_small_freq}
\hat{D}_y^\text{eq}(s) \simeq \frac{D_y^0}{s}[1 - n(\pi-1)] - \frac{D_y^0}{2} n \pi \tau \ln(s\tau).
\end{align}
The stationary-state diffusion coefficient $D_y^\text{eq} := D_y^\text{eq}(t\to \infty)$ is obtained from the frequency-domain 
representation via
\begin{align}
D_y^\text{eq} = \lim_{s\to 0} s\hat{D}_y^\text{eq}(s) = D_y^0[1 - n(\pi-1)].
\end{align}
The logarithmic divergence for small frequencies~[Eq.~\eqref{eq:diff_coeff_eq_small_freq}] arises due to the repeated
encounters of the tracer with the same obstacle and corresponds to an algebraic decay to the stationary-state diffusion
coefficient in the time domain:
\begin{align}
D_y^\text{eq}(t) - D_y^\text{eq} \simeq \frac{n\pi D_y^0}{2} \frac{\tau}{t},\quad t\to\infty.
\end{align}
In particular, we recover the algebraic decay for the long-time behavior of the velocity-autocorrelation function of the
tracer~\cite{Nieuwenhuizen:PRL_57:1986}:
\begin{align}
\frac{\diff}{\diff t} D_y^\text{eq}(t) \simeq -\frac{n\pi D_y^0}{2}\frac{\tau}{t^2},\quad t\to\infty , 
\end{align}
with bare perpendicular diffusion coefficient $D_y^0 = 1/4\tau$.

\begin{figure}
\includegraphics{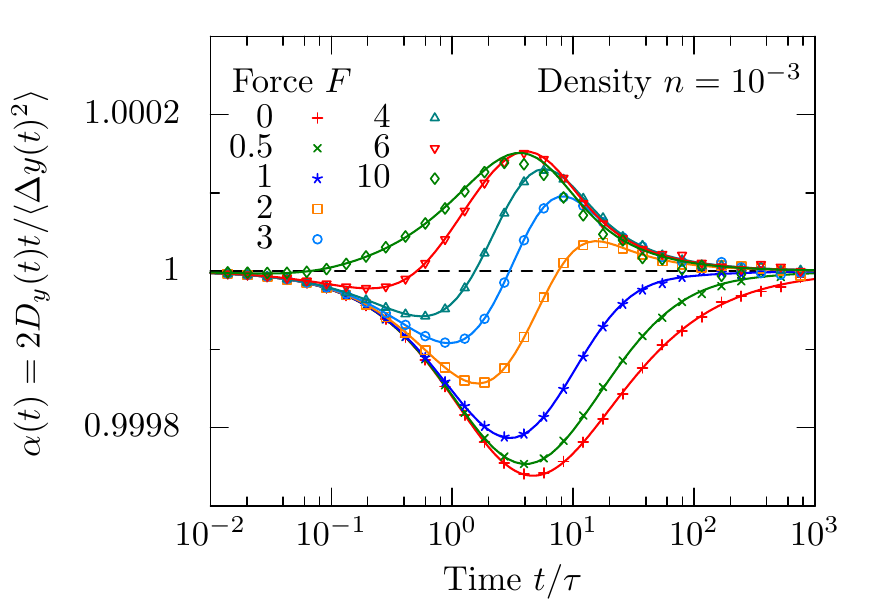}
\caption{\label{fig:subsuperdiffy}
Local exponent $\alpha(t) = \diff\ln(\langle \Delta y(t)^2\rangle)/\diff\!\ln(t)$ measuring
subdiffusive $\alpha <1$ and superdiffusive $\alpha >1$ behavior perpendicular to the applied force
on the tracer. Solid lines correspond to the theory, and symbols represent data from the stochastic
simulation.
}
\end{figure}

In the presence of a force $F > 0$, the time-dependent behavior of the diffusion coefficient becomes much more complex
as shown in Fig.~\ref{fig:diffcoeffy}, where we compare stochastic simulations (see
Appendix~\ref{Appendix:Stochastic_simulation}) with the numerically inverted analytic solution (see
Appendix~\ref{Appendix:Numerical_Inversion}).  For any finite force $F$, the approach to the stationary diffusion
coefficient perpendicular to the force, $D_y := D_y(t\to\infty)$, is always nonmonotonic.  First, the diffusivity of the
tracer decreases, similar to the equilibrium case, until a point of least diffusivity is reached at intermediate times.
Then, the diffusion coefficient increases again until the stationary state is reached. For sufficiently large driving,
the stationary diffusion coefficient $D_y$ becomes larger than the short-time diffusion coefficient in the presence of
obstacles, $D_y(t\to 0) = D_y^0(1-n)$. The perpendicular diffusion coefficient is bounded by $D_y(F\to\infty) = D_y^0(1
- n/2)$ in first order of the density, as can be rationalized by the analytic solution~[Fig.~\ref{fig:stat_diffcoeffy}].
Deviations from the first-order theory become apparent at large forces ($F=10$)~[Fig.~\ref{fig:diffcoeffy}]. This
observation is consistent with the general insight, that the range of validity of the first-order solution depends on
the magnitude of the force: The larger the force, the smaller the density has to be for the theory to be an accurate
representation of the simulation~\cite{Leitmann:PRL_111:2013,Leitmann:PRL_118:2017}.
 
For small driving $F \to 0$, a series expansion of our solution reveals that the stationary diffusion coefficient
perpendicular to the field exhibits a density-induced nonanalytic behavior: 
\begin{align} \label{eq:diffcoeff_y_small_driving}
D_y = D_y^\text{eq} + n D_y^0 F^2[A \ln(1/|F|) + B] + \mathcal{O}[F^2\ln(1/|F|)]^2
\end{align}
with coefficients $A = (\pi+4)/16 \approx 0.446$ and $B = [7(\pi+4)\ln(2) - \pi(\pi^2 - \pi + 2)/(\pi-2)]/32 \approx
0.332$. Similar nonanalytic behavior has been found for the stationary velocity and the stationary diffusion coefficient
parallel to the force in the driven lattice Lorentz gas~\cite{Leitmann:PRL_111:2013,Leitmann:PRL_118:2017}. 

The question of how the increase of the diffusivity is related to a transient superdiffusive behavior of the tracer can be answered by 
considering the local exponent 
\begin{align}
\alpha(t) := \frac{\diff \ln\bigl(\langle \Delta y(t)^2\rangle\bigr)}{\diff\! \ln(t)} 
= \frac{2 D_y(t) t}{\langle \Delta y(t)^2\rangle}. 
\end{align}
Then, (transient) subdiffusive behavior is defined by $\alpha(t) < 1$ while superdiffusion is related to
$\alpha(t) > 1$. 
For forces $F \gtrsim 1$, the increase in the diffusivity corresponds to a superdiffusive increase of the fluctuations
[Fig.~\ref{fig:subsuperdiffy}]. For increasing force, the time window of the superdiffusive regime becomes more and
more pronounced, and for the highest force considered ($F = 10$), the stationary
state is essentially approached only superdiffusively.

\subsection{Non-Skellam parameter}

\begin{figure}
\includegraphics{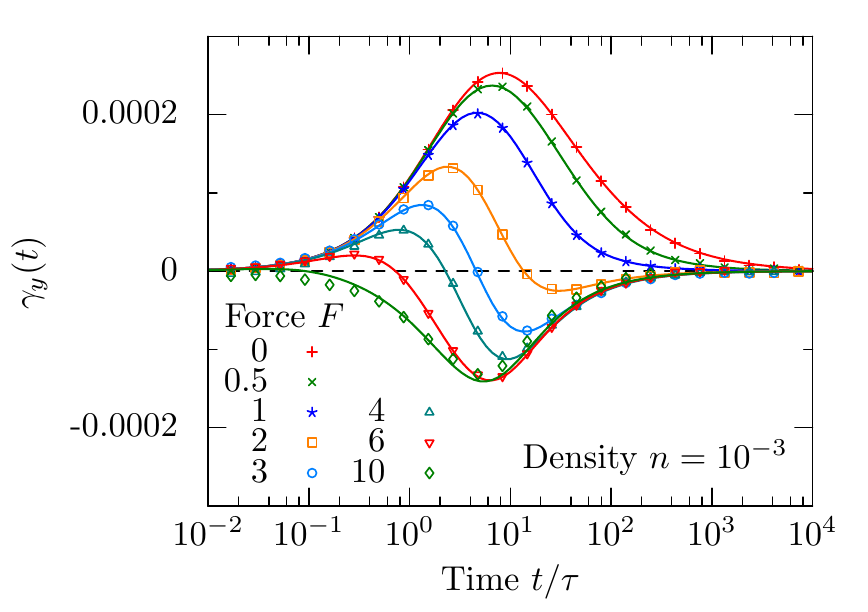}
\caption{ \label{fig:non_skellam_parameter}
Non-Skellam parameter $\gamma_y(t)$ [Eq.~\eqref{eq:non_skellam_parameter}] for different
strength of the force $F$. Solid lines correspond to the analytic solution. Symbols represent results from stochastic
simulations.
}
\end{figure}

To further assess the effects of the obstacle disorder on the motion of the tracer,
we evaluate the mean-quartic displacement perpendicular to the force, $\langle \Delta y(t)^4 \rangle$,
and define the non-Skellam parameter:
\begin{align} \label{eq:non_skellam_parameter}
\gamma_y(t) = \frac{\langle \Delta y(t)^4\rangle - \langle \Delta y(t)^2\rangle}{3\langle \Delta y(t)^2\rangle^2} - 1.
\end{align}
The choice of the non-Skellam parameter as an indicator of obstacle-induced effects is motivated by the fact that it vanishes for all times in the absence of obstacles.
This can be inferred from the time-dependent behavior of the bare mean-quartic displacement:
\begin{align}
\langle \Delta y(t)^4\rangle_0 = \frac{\partial^4}{\partial k_y^4} F_0(\vec{k},t)\Bigl|_{\vec{k}=0} = \langle \Delta y(t)^2\rangle_0 + 3\langle \Delta y(t)^2\rangle_0^2 .
\end{align}
In the presence of obstacles, the non-Skellam parameter shows deviations from the obstacle-free
case~[Fig.~\ref{fig:non_skellam_parameter}]. For small forces, the deviations are positive until negative contributions
emerge for increasing driving. Interestingly, the transformed local exponent $1 - \alpha(t)$ strongly resembles the
non-Skellam parameter such that negative/positive contributions in the non-Skellam parameter correspond to
superdiffusive/subdiffusive behavior in the local exponent.

\begin{figure}
\includegraphics{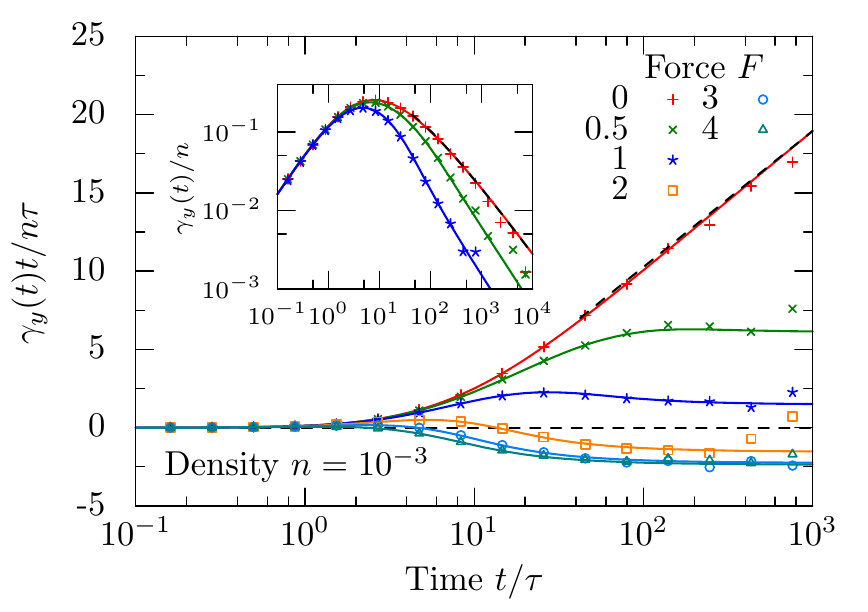}
\caption{ \label{fig:non_skellam_parameter_log}
Non-Skellam parameter $\gamma_y(t)$ [Eq.~\eqref{eq:non_skellam_parameter}] normalized by the density $n$ for different
strength of the force $F$ and the equilibrium case, $F = 0$. Solid lines correspond to the analytic solution, and symbols
represent results from stochastic simulations. The dashed line represents the long-time behavior of the non-Skellam
parameter in equilibrium~[Eq.~\eqref{eq:non_skellam_parameter_longtime}].
}
\end{figure}

In equilibrium, the long-time behavior of the non-Skellam parameter 
can be elaborated via the low-frequency expansion, revealing slow algebraic tails:
\begin{align} \label{eq:non_skellam_parameter_longtime}
\gamma_y(t)/n =  A_\gamma \frac{\ln(t/\tau)}{t/\tau} + \frac{B_\gamma}{t/\tau} +
\mathcal{O}[\ln(t)/t]^2 , \quad t\to\infty,
\end{align}
with coefficients $A_\gamma = \pi+2/\pi \approx 3.778$ and 
$B_\gamma = -(\pi^4 + \pi^3 - 7\pi^2 + 2\pi-4)/(\pi-2)\pi + A_\gamma[\gamma_\text{e} + 3\ln(2)] \approx -7.142$ with Euler's
constant $\gamma_\text{e} \approx 0.577$.
In equilibrium, we observe the leading decay $\ln(t)/t$ due to persistent correlations, whereas for any finite driving, one
recovers a $t^{-1}$ tail as anticipated from the central limit theorem for weakly correlated
increments~[Fig.~\ref{fig:non_skellam_parameter_log}].

The logarithmic contribution in the non-Skellam parameter for long times manifests itself as a density-induced
logarithmic divergence of the super-Burnett coefficient in
equilibrium~\cite{Ernst:JSP_26:1981,vanBeijeren:RMP_54:1982,Nieuwenhuizen:PA_157:1989}:
\begin{align}
\begin{split}
D^{(2)}(t) &= \frac{1}{4!}\frac{\diff}{\diff t}\Bigl[\langle \Delta y(t)^4\rangle - 3\langle \Delta y(t)^2\rangle^2 \Bigr] \\
&= \frac{n}{32\tau}\Bigl(\pi+\frac{2}{\pi}\Bigr)\ln(t/\tau) + \mathcal{O}(1), \quad t\to\infty.
\end{split}
\end{align}

It is also possible to define a non-Skellam parameter $\gamma_x(t)$ parallel to the applied force in terms of the
centralized fluctuating variable for the displacement, $\Delta x(t) - \langle \Delta x(t)\rangle$. The analytic solution
can be worked out in principle but is considerably more involved since also the mean displacement $\langle \Delta
x(t)\rangle$ and the mean-cubic displacement $\langle \Delta x(t)^3\rangle$ contribute.  Here, we only give a
characteristic feature of the non-Skellam parameter along the force, $\gamma_x(t)$, by employing an asymptotic model for
large forces~\cite{Leitmann:PRL_118:2017}. In this model, the tracer always jumps along the field and moves along one-dimensional lanes until it hits
an obstacle and stops. For large forces, this model captures the dynamics in the driven
lattice Lorentz gas as long as the diffusive time scale is not reached such that the tracer passes the obstacle blocking
its path~\cite{Leitmann:PRL_118:2017}.  Then, by working out the solution, the non-Skellam parameter shows an increase of $\gamma_x(t) = n
e^{3F/2}(t/\tau)^3/960$ at intermediate times which can also be observed in stochastic simulation at large driving and
small obstacle densities. A similar behavior is also found in the fluctuations along the field measured in terms of the
variance $\text{Var}_x(t) = \langle \Delta x^2(t)\rangle - \langle \Delta x(t)\rangle^2$. For large forces, the variance
increases superdiffusively, $\text{Var}_x(t)\sim t^3$, and the respective window of superdiffusive motion grows
exponentially with the force $\sim e^{F/2}$~\cite{Leitmann:PRL_118:2017}. Thus, since the window of superdiffusion can
be made arbitrarily large, the driven lattice Lorentz gas exhibits a true superdiffusive exponent of $3$ parallel to the
force that is again found in the non-Skellam parameter at intermediate times, $\gamma_x(t) \sim t^3$.

\section{Summary and Conclusion} \label{sec:summary_conclusion}

We have solved for the dynamics of a tracer particle on a planar lattice in the presence of immobile, hard obstacles.
At time zero, a force pulling the tracer along a lattice direction is switched on, driving the system out of equilibrium.
Here, the dynamics of the tracer has been analyzed in terms of the fluctuations perpendicular to the applied force as encoded in the 
diffusion coefficient, the local exponent, and the non-Skellam parameter.

In equilibrium where no force acts on the tracer, the time-dependent diffusion coefficient decreases monotonically
to its stationary state value. This behavior is no longer true for nonvanishing driving, and the point of least
diffusivity is always attained at intermediate times. For increasing force, the stationary diffusion coefficient
perpendicular to the force can increase beyond the short-time diffusion coefficient in the presence of obstacles, but it never
grows beyond a certain bound irrespective of the force.

The time-dependent local exponent shows subdiffusive and superdiffusive transient behavior of the order of the density
when the force exceeds a certain threshold. A similar behavior is found for the non-Skellam parameter, which encodes the
effect on the dynamics induced by the obstacles such that positive (negative) contributions in the non-Skellam parameter
indicate subdiffusive (superdiffusive) behavior as found in the local exponent.  In equilibrium, the non-Skellam
parameter exhibits a logarithmic dependence for long times, which is also reflected in a logarithmic divergence of the
respective super-Burnett coefficient.

The fluctuations parallel to the applied force have been evaluated just recently~\cite{Leitmann:PRL_118:2017}. For small
driving, both directions exhibit similar behavior in terms of density-induced nonanalytic contributions in the
stationary-state diffusion coefficients and the qualitative behavior of the time-dependent approach to the stationary
state. In contrast to the stationary-state diffusion coefficient perpendicular to the field where the effects are of the
order of the density and bounded from above, the diffusion coefficient parallel to the field exhibits an exponential
growth for increasing driving and can become arbitrarily large. Moreover, the local exponent exhibits a true
superdiffusive exponent of $3$ for large driving, which is absent in the fluctuations perpendicular to the field. This
exponent is again found at intermediate times in the non-Skellam parameter parallel to the field. 

The saturation of the perpendicular diffusion coefficient for increasing driving is a peculiarity of the lattice and is
not observed in continuum. There, a probe particle performs Brownian motion in the presence of other bath particles, and the
diffusion coefficient perpendicular to the applied field was evaluated earlier and increases linearly with the
driving for strong forces~\cite{Zia:JFM_658:2010}. The differences can be attributed to the particular realization of the model on a
lattice, as also indicated by the difference in the force dependence of the parallel diffusion coefficient for large
driving~\cite{Leitmann:PRL_118:2017}.

For the driven lattice Lorentz gas, the intermediate scattering function is known in the frequency domain in first order
of the density of obstacles and for arbitrarily strong driving. Thus, it is possible in principle to evaluate the
probability distribution of the tracer displacements from the analytic solution. In particular, this allows studying the
tails of the probability distribution encoding the rare events where the motion is anticipated to differ drastically
from the central limit theorem. 

The solution strategy elaborated here can be extended to the three-dimensional case. There, one has to solve for a
$7\times 7$ matrix problem consisting of the space spanned by the obstacle and the six nearest neighbors. Again, it is
advantageous to exploit the symmetries, and one can convince oneself that the problem reduces to solving for $3\times 3$
matrices. The respective propagators in three dimensions are more complicated but can still be expressed in terms of
elliptic integrals~\cite{Joyce:JPAMG_34:2001,Joyce:JPAMG_35:2002} such that the overall numerical evaluation of the transport
properties should still be feasible.

The nonanalytic dependence on the force as well as the vanishing of the long-time tails should prevail for all densities
except in the vicinity of the percolation transition. Both are directly connected to each other in terms of the
propagators and therefore they are merely two sides of the same coin. Upon approaching the percolation transition where the
infinite cluster becomes self-similar and anomalous transport emerges in
equilibrium~\cite{Stauffer:Taylor:1994,benAvraham:CUP:2000,Kammerer:EPL_84:2008}, driving may introduce new interesting
phenomena~\cite{vanBeijeren:PRL_54:1985,Yau:AoM_159:2004}, such as an anomalous mean displacement~\cite{Bouchaud:PR_195:1990}.

\begin{acknowledgments}
We gratefully acknowledge support by the DFG research unit FOR1394 'Nonlinear response to probe vitrification'.
\end{acknowledgments}

\appendix
\section{Stochastic simulation} \label{Appendix:Stochastic_simulation}

The stochastic simulation of the driven lattice Lorentz gas is performed in discrete time measured in the number of jumps $J$
of the tracer particle. The moments of the displacement in discrete time, $\langle \Delta y^m_J \rangle$, are
then transformed to continuous time via a Poisson transform~\cite{Haus:PR_150:1987,Frenkel:PLA_121:1987}
\begin{align}
\langle \Delta y(t)^m\rangle = \sum_{J = 0}^\infty \langle \Delta y^m_J \rangle \Psi_J(\Gamma t),
\end{align}
with $\Psi_J(t) = (t/\tau)^J e^{-t/\tau}/J!$ and $\Gamma = [\cosh(F/2)+1]/2$.

For small obstacle densities, the fluctuations of the free dynamics are much larger than the obstacle-induced response,
and it is advantageous to adapt the approach of Ref.~\cite{Frenkel:PLA_121:1987}.
Following this idea, the displacement $\Delta y_J$ is split into two distinct parts:
\begin{align}
\Delta y_J = \Delta y_{0,J} + (\Delta y_J - \Delta y_{0,J}).
\end{align}
The first contribution, $\Delta y_{0,J}$, represents the free dynamics in the absence of obstacles, whereas the second
contribution, $\Delta y_J - \Delta y_{0,J}$, accounts for the influence of the obstacles on the dynamics of the tracer.
By measuring the second contribution in simulations, the mean-square displacement perpendicular to the field is then obtained
as
\begin{align}
\langle \Delta y(t)^2\rangle = \langle \Delta y(t)^2\rangle_0 + \sum_{J=0}^\infty\langle\Delta y_J^2 - \Delta
y_{0,J}^2\rangle \Psi_J(\Gamma t), 
\end{align}
with the mean-square displacement in the absence of obstacles, $\langle \Delta y(t)^2\rangle_0 = 2 D_y^0 t$.

\section{Numerical inversion of the frequency-dependent response functions} \label{Appendix:Numerical_Inversion}

The relevant response functions are obtained in the frequency domain and are transformed to the time domain via an
inverse Laplace transform. To illustrate the technique, we consider a real function $f(t)$ and apply the
Laplace transform
\begin{align}
\hat{f}(s) = \int_0^\infty\diff t\ e^{-st} f(t),
\end{align}
with complex frequency $s=\sigma+\img\omega$ in the complex right half-plane. We take the real part of both sides and
express the real part of $\hat{f}(s)$ as twice the cosine transform of the symmetric function $e^{-\sigma|t|}f(|t|)$:
\begin{align}
2\text{Re}[\hat{f}(\sigma+\img\omega)] = \int_{-\infty}^\infty \diff t\ e^{-\sigma |t|}f(|t|) \cos(\omega t) .
\end{align}
With the relation
\begin{align}
\frac{1}{\pi}\int_{-\infty}^\infty \diff\omega\ \cos(\omega t)\cos(\omega t') = \delta(t+t') + \delta(t-t'),
\end{align}
which can derived from the formal representation of the Dirac delta function
${2\pi\delta(t - t')} = \int_{-\infty}^\infty \diff \omega\ e^{\img\omega(t-t')}$~\cite{Olver:2010:NHMF,NIST:DLMF},
we obtain the real function $f(t)$ from the Laplace transform $\hat{f}(s)$ via 
\begin{align} \label{eq:fourier_inversion}
f(t) = \frac{2 e^{\sigma t}}{\pi} \int_0^\infty \diff \omega\ \text{Re}[\hat{f}(\sigma + \img \omega)]\cos(\omega t), \quad t \geq 0.
\end{align}
The back-transform [Eq.~\eqref{eq:fourier_inversion}] is implemented numerically by a suitable Filon
formula~\cite{Tuck:MoC_21:1967}. For the numerical evaluation of the complete elliptic integrals of the first and second
kind, we use an implementation provided by the mpmath multi-precision library~\cite{Carlson:NA_10:1995,mpmath}.
In general, it is numerically much more stable to use $\sigma = 0$, since the exponential increase vanishes. However,
this is not possible for functions with finite long-time limit $f(t\to\infty) = \lim_{s\to 0} s\hat{f}(s)$  such as
the time-dependent diffusion coefficient $D_y(t)$. In these cases, it is advantageous to perform the numerical
back-transform on the function $\Delta \hat{f}(s) = \hat{f}(s) - f(t\to\infty)/s$ for $\sigma = 0$ with
\begin{align} \label{eq:fourier_inversion_modified}
f(t) - f(t\to\infty) = \frac{2}{\pi} \int_0^\infty \diff \omega\ \text{Re}[\Delta \hat{f}(\img \omega)]\cos(\omega t),
\quad t \geq 0,
\end{align}
and trivially add the long-time limit $f(t\to\infty)$ after the numerical back-transform.

\section{Symmetry transformation} \label{Appendix:Group_Theory}
In the case of no driving, $F=0$, the dihedral group $D_4$ contains all symmetry transformations that are possible for
the obstacle site and its four neighbors. A matrix representation of the group can be directly obtained by writing the
symmetry transformations with respect to the basis $\{\vec{e}_{-2},\vec{e}_{-1},\vec{e}_0,\vec{e}_1,\vec{e}_2\}$
introduced earlier. For example, the
counterclockwise rotation $C_4$ by an angle of $\pi/2$ is given by
\begin{align}
D(C_4) = 
\begin{pmatrix}
0 & 1 & 0 & 0 & 0\\
0 & 0 & 0 & 0 & 1\\
0 & 0 & 1 & 0 & 0\\
1 & 0 & 0 & 0 & 0\\
0 & 0 & 0 & 1 & 0
\end{pmatrix}.
\end{align}


The dihedral group $D_4$ consists of five different classes $\mathcal{C}_k$ and thus five irreducible
representations $\alpha$ with characters $\chi^\alpha(\mathcal{C}_k)$. 
The number $m_\alpha$ of irreducible representations $\alpha$ in our matrix representation can then be determined by the
formula~\cite{Tinkham:Dover:2003}
\begin{align}
m_\alpha = \frac{1}{h} \sum_k N_k [\chi^\alpha(\mathcal{C}_k)]^* \chi(\mathcal{C}_k),
\end{align}
with the number $N_k$ of elements in the class $\mathcal{C}_k$, $\chi(\mathcal{C}_k)$ the character of the matrix
representation, and $h$ the number of elements in the group $D_4$. Then, one obtains $m_{A_1} = 2$, $m_{A_2} = 0$,
$m_{B_1} = 1$, $m_{B_2} = 0$, and $m_E = 1$, and the respective projection operators $P^\alpha$ are then determined by  
\begin{align}
P^\alpha = \frac{l_\alpha}{h}\sum_g[\chi^\alpha(g)]^*D(g)
\end{align}
as a sum over all group elements $g$ and the dimensionality $l_\alpha$ of the irreducible representation $\alpha$. 

This leads to the projection operators 
\begin{align}
P^{A_1} = \frac{1}{4}
\begin{pmatrix}
\phantom{-}1 & \phantom{-}1 & \phantom{-}0 & \phantom{-}1 & \phantom{-}1\\
\phantom{-}1 & \phantom{-}1 & \phantom{-}0 & \phantom{-}1 & \phantom{-}1\\
\phantom{-}0 & \phantom{-}0 & \phantom{-}4 & \phantom{-}0 & \phantom{-}0\\
\phantom{-}1 & \phantom{-}1 & \phantom{-}0 & \phantom{-}1 & \phantom{-}1\\
\phantom{-}1 & \phantom{-}1 & \phantom{-}0 & \phantom{-}1 & \phantom{-}1
\end{pmatrix}, \\
P^{B_1} = \frac{1}{4}
\begin{pmatrix}
\phantom{-}1 & -1 & \phantom{-}0 & -1 & \phantom{-}1\\
-1 & \phantom{-}1 & \phantom{-}0 & \phantom{-}1 & -1\\
\phantom{-}0 & \phantom{-}0 & \phantom{-}0 & \phantom{-}0 & \phantom{-}0\\
-1 & \phantom{-}1 & \phantom{-}0 & \phantom{-}1 & -1\\
\phantom{-}1 & -1 & \phantom{-}0 & -1 & \phantom{-}1
\end{pmatrix}, \\
P^{E} = \frac{1}{2}
\begin{pmatrix}
\phantom{-}1 & \phantom{-}0 & \phantom{-}0 & \phantom{-}0 & -1\\
\phantom{-}0 & \phantom{-}1 & \phantom{-}0 & -1 & \phantom{-}0\\
\phantom{-}0 & \phantom{-}0 & \phantom{-}0 & \phantom{-}0 & \phantom{-}0\\
\phantom{-}0 & -1 & \phantom{-}0 & \phantom{-}1 & \phantom{-}0\\
-1 & \phantom{-}0 & \phantom{-}0 & \phantom{-}0 & \phantom{-}1
\end{pmatrix}, 
\end{align}
from which the orthogonal matrix $M$~[Eq.~\eqref{eq:orth_trans_matrix}] is generated.

\onecolumngrid
\section{Matrix elements of the scattering $t$-matrix} \label{appendix:explicit_expressions_matrix_elements}
Here, we give explicit expressions for the matrix elements of the scattering $t$-matrix in terms of the
propagators $g_{00}$, $g_{10}$, $g_{11}$, and $g_{20}$. First, we give the determinant of $(1-v'G_0')$:
\begin{align}
\begin{split}
\text{det}[(1-v'G_0')] &= 
(s/4) [1+(1/4\tau) (g_{20}-g_{00})] \biglb[2 g_{11} \biglb(s \{2 g_{10}-g_{00} [1+(1/4\tau)(g_{20}-g_{00})]\} + 8 (1/4\tau) g_{10}+1\bigrb) + 
\\ &\phantom{==} -s \{g_{00} [-(1/4\tau) g_{20}^2+ (1/4\tau) g_{00}^2-3g_{00} + g_{20}]+g_{10} (g_{20}+3 g_{00})\} -8 (1/4\tau) g_{11}^2 + 
\\ &\phantom{==} +8 (1/4\tau) g_{00} (g_{00}-2 g_{10})-2 g_{10} \cosh (F/2)-2 g_{10}+g_{20}+g_{00}\bigrb].
\end{split}
\end{align}
The matrix elements $\langle p_y |\hat{t}| p_y\rangle$, $\langle p_x |\hat{t}| n\rangle$, $\langle d_{xy}|\hat{t}|
n \rangle$, and $\langle s |\hat{t}| n\rangle$ can then be written in the following way:
\begin{align}
\begin{split}
\langle p_y | \hat{t} | p_y \rangle &= 
(1/4\tau)[1+(1/4\tau) (g_{20}-g_{00})]^{-1}, 
\end{split} \\ \nonumber \\
\begin{split}
\text{det}[(1-v'G_0')]\cdot\langle p_x | \hat{t} | n \rangle &= 
4 \img (1/4\tau) s [1+(1/4\tau) (g_{20}-g_{00})] \sinh (F/2) \{(1/4\tau) [2 g_{10}^2-2 g_{11} g_{10}+(g_{20}+g_{00}) g_{10} + 
\\ &\phantom{==} -g_{00} (g_{20}+g_{00})] \cosh (F/2) +(1/4\tau) [2 g_{10}^2-g_{00} (g_{20}+g_{00})]+g_{00}\} , 
\end{split} \\ \nonumber \\
\begin{split}
\text{det}[(1-v'G_0')]\cdot\langle d_{xy} | \hat{t} | n \rangle &= 
(2/\sqrt{5}) (1/4\tau)^2 (-2 g_{11}+g_{20}+g_{00}) s [1 + (1/4\tau) (g_{20}-g_{00})] \times 
\\ &\phantom{==}\times \{(1/4\tau) [g_{00} (2 g_{11}+g_{20}+g_{00})-4 g_{10}^2]-g_{00}\} \sinh^2(F/2) ,
\end{split} \\ \nonumber \\
\begin{split}
\text{det}[(1-v'G_0')]\cdot\langle s | \hat{t} | n \rangle &= 
(2/5) (1/4\tau) s [1 + (1/4\tau) (2 g_{11}-g_{20}-g_{00})] [1 + (1/4\tau) (g_{20}-g_{00})]\times
\\ &\phantom{==}\times [-4 (1/4\tau) g_{10}^2+(1/4\tau) g_{00} (2 g_{11}+g_{20}+g_{00})+4 g_{10}] \sinh ^2(F/2) .
\end{split}
\end{align}
For the last three entries, we have multiplied by the determinant $\text{det}[(1-v'G_0')]$ to simplify the resulting
expressions. The matrix element $\langle p_x | \hat{t} | p_x\rangle$ with
\begin{align}
\text{det}[(1-v'G_0')]\cdot \langle p_x | \hat{t} | p_x \rangle = C_{22} \langle p_x |\hat{v}|p_x \rangle +
C_{32} \langle d_{xy}|\hat{v}|p_x \rangle + C_{42} \langle s |\hat{v}|p_x \rangle ,
\end{align}
is expressed in terms of the cofactors
\begin{align}
\begin{split}
C_{22} &= 
s [1+(1/4 \tau) (g_{20}-g_{00})] \biglb[(1/4 \tau) \biglb(g_{10}^2 \{(1/4 \tau) [8 g_{11}-4 (g_{20}+g_{00})]+2\}+ \\
&\phantom{==}+ g_{00} \{(1/4 \tau) [(g_{20}+g_{00})^2-4 g_{11}^2]-g_{20}-g_{00}\}\bigrb) \cosh (F/2) +(1/4 \tau) [2
g_{10}^2-g_{00} (g_{20}+g_{00})]+g_{00}\bigrb] , 
\end{split} \\ \nonumber \\
\begin{split}
C_{32} &= 
(1/4\sqrt{2}\tau) (g_{20}-g_{00}) s [1 + (1/4 \tau) (g_{20}-g_{00})] \{(1/4 \tau) [4 g_{10}^2-g_{00} (2
g_{11}+g_{20}+g_{00})]+g_{00}\} \sinh (F/2) , 
\end{split} \\ \nonumber \\
\begin{split}
C_{42} &= 
(1/4\sqrt{10}\tau) [1 + (1/4\tau) (g_{20}-g_{00})] \sinh (F/2) \biglb[-4 (1/4\tau)^2 (g_{20}-g_{00}) [2 g_{11} (g_{00}-g_{10})+g_{10} (-2 g_{10}+g_{20}+g_{00})] + 
\\ &\phantom{==}-8 (1/4\tau) \biglb(g_{10}^2 [4 (1/4\tau) g_{11}-(1/4\tau) g_{20}-3 (1/4\tau) g_{00}+1]+g_{00} \{(1/4\tau) [g_{00} (g_{20}+g_{00})-2 g_{11}^2]-g_{00}\}\bigrb) \cosh (F/2) +
\\ &\phantom{==}+(g_{20}-g_{00}) g_{00} s [(1/4\tau) (-2 g_{11}+g_{20}+g_{00})-1]-4 (1/4\tau) (2 g_{10}-g_{20}-g_{00})
(g_{10}+g_{00})-4 g_{00}\bigrb] ,
\end{split}
\end{align}
and the matrix elements of the obstacle potential $\hat{v}$ in the basis $\{p_y, p_x, d_{xy}, s, n\}$ with $\langle
p_x|\hat{v}|p_x \rangle = \cosh(F/2)/4\tau$, $\langle d_{xy}|\hat{v}|p_x \rangle = -\sinh(F/2)/4\sqrt{2}\tau$, and
$\langle s |\hat{v}|p_x \rangle = \sqrt{5/2}\sinh(F/2)/4\tau$.

\twocolumngrid


\begin{thebibliography}{66}%
\makeatletter
\providecommand \@ifxundefined [1]{%
 \@ifx{#1\undefined}
}%
\providecommand \@ifnum [1]{%
 \ifnum #1\expandafter \@firstoftwo
 \else \expandafter \@secondoftwo
 \fi
}%
\providecommand \@ifx [1]{%
 \ifx #1\expandafter \@firstoftwo
 \else \expandafter \@secondoftwo
 \fi
}%
\providecommand \natexlab [1]{#1}%
\providecommand \enquote  [1]{``#1''}%
\providecommand \bibnamefont  [1]{#1}%
\providecommand \bibfnamefont [1]{#1}%
\providecommand \citenamefont [1]{#1}%
\providecommand \href@noop [0]{\@secondoftwo}%
\providecommand \href [0]{\begingroup \@sanitize@url \@href}%
\providecommand \@href[1]{\@@startlink{#1}\@@href}%
\providecommand \@@href[1]{\endgroup#1\@@endlink}%
\providecommand \@sanitize@url [0]{\catcode `\\12\catcode `\$12\catcode
  `\&12\catcode `\#12\catcode `\^12\catcode `\_12\catcode `\%12\relax}%
\providecommand \@@startlink[1]{}%
\providecommand \@@endlink[0]{}%
\providecommand \url  [0]{\begingroup\@sanitize@url \@url }%
\providecommand \@url [1]{\endgroup\@href {#1}{\urlprefix }}%
\providecommand \urlprefix  [0]{URL }%
\providecommand \Eprint [0]{\href }%
\providecommand \doibase [0]{http://dx.doi.org/}%
\providecommand \selectlanguage [0]{\@gobble}%
\providecommand \bibinfo  [0]{\@secondoftwo}%
\providecommand \bibfield  [0]{\@secondoftwo}%
\providecommand \translation [1]{[#1]}%
\providecommand \BibitemOpen [0]{}%
\providecommand \bibitemStop [0]{}%
\providecommand \bibitemNoStop [0]{.\EOS\space}%
\providecommand \EOS [0]{\spacefactor3000\relax}%
\providecommand \BibitemShut  [1]{\csname bibitem#1\endcsname}%
\let\auto@bib@innerbib\@empty
\bibitem [{\citenamefont {Squires}(2008)}]{Squires:LM_24:2008}%
  \BibitemOpen
  \bibfield  {author} {\bibinfo {author} {\bibfnamefont {T.~M.}\ \bibnamefont
  {Squires}},\ }\href {\doibase 10.1021/la7023692} {\bibfield  {journal}
  {\bibinfo  {journal} {Langmuir}\ }\textbf {\bibinfo {volume} {24}},\ \bibinfo
  {pages} {1147} (\bibinfo {year} {2008})}\BibitemShut {NoStop}%
\bibitem [{\citenamefont {Wilson}\ and\ \citenamefont
  {Poon}(2011)}]{Wilson:PCCP_13:2011}%
  \BibitemOpen
  \bibfield  {author} {\bibinfo {author} {\bibfnamefont {L.~G.}\ \bibnamefont
  {Wilson}}\ and\ \bibinfo {author} {\bibfnamefont {W.~C.~K.}\ \bibnamefont
  {Poon}},\ }\href {\doibase 10.1039/C0CP01564D} {\bibfield  {journal}
  {\bibinfo  {journal} {Phys. Chem. Chem. Phys.}\ }\textbf {\bibinfo {volume}
  {13}},\ \bibinfo {pages} {10617} (\bibinfo {year} {2011})}\BibitemShut
  {NoStop}%
\bibitem [{\citenamefont {Puertas}\ and\ \citenamefont
  {Voigtmann}(2014)}]{Puertas:JPCM_26:2014}%
  \BibitemOpen
  \bibfield  {author} {\bibinfo {author} {\bibfnamefont {A.~M.}\ \bibnamefont
  {Puertas}}\ and\ \bibinfo {author} {\bibfnamefont {{\relax Th}.}~\bibnamefont
  {Voigtmann}},\ }\href {\doibase 10.1088/0953-8984/26/24/243101} {\bibfield
  {journal} {\bibinfo  {journal} {J. Phys. Condens. Matter}\ }\textbf {\bibinfo
  {volume} {26}},\ \bibinfo {pages} {243101} (\bibinfo {year}
  {2014})}\BibitemShut {NoStop}%
\bibitem [{\citenamefont {Mason}\ and\ \citenamefont
  {Weitz}(1995)}]{Mason:PRL_74:1995}%
  \BibitemOpen
  \bibfield  {author} {\bibinfo {author} {\bibfnamefont {T.~G.}\ \bibnamefont
  {Mason}}\ and\ \bibinfo {author} {\bibfnamefont {D.~A.}\ \bibnamefont
  {Weitz}},\ }\href {\doibase 10.1103/PhysRevLett.74.1250} {\bibfield
  {journal} {\bibinfo  {journal} {Phys. Rev. Lett.}\ }\textbf {\bibinfo
  {volume} {74}},\ \bibinfo {pages} {1250} (\bibinfo {year}
  {1995})}\BibitemShut {NoStop}%
\bibitem [{\citenamefont {Habdas}\ \emph {et~al.}(2004)\citenamefont {Habdas},
  \citenamefont {Schaar}, \citenamefont {Levitt},\ and\ \citenamefont
  {Weeks}}]{Habdas:EPL_67:2004}%
  \BibitemOpen
  \bibfield  {author} {\bibinfo {author} {\bibfnamefont {P.}~\bibnamefont
  {Habdas}}, \bibinfo {author} {\bibfnamefont {D.}~\bibnamefont {Schaar}},
  \bibinfo {author} {\bibfnamefont {A.~C.}\ \bibnamefont {Levitt}}, \ and\
  \bibinfo {author} {\bibfnamefont {E.~R.}\ \bibnamefont {Weeks}},\ }\href
  {\doibase 10.1209/epl/i2004-10075-y} {\bibfield  {journal} {\bibinfo
  {journal} {Europhys. Lett.}\ }\textbf {\bibinfo {volume} {67}},\ \bibinfo
  {pages} {477} (\bibinfo {year} {2004})}\BibitemShut {NoStop}%
\bibitem [{\citenamefont {Carpen}\ and\ \citenamefont
  {Brady}(2005)}]{Carpen:JRHEO_49:2005}%
  \BibitemOpen
  \bibfield  {author} {\bibinfo {author} {\bibfnamefont {I.~C.}\ \bibnamefont
  {Carpen}}\ and\ \bibinfo {author} {\bibfnamefont {J.~F.}\ \bibnamefont
  {Brady}},\ }\href {\doibase 10.1122/1.2085174} {\bibfield  {journal}
  {\bibinfo  {journal} {J. Rheol.}\ }\textbf {\bibinfo {volume} {49}},\
  \bibinfo {pages} {1483} (\bibinfo {year} {2005})}\BibitemShut {NoStop}%
\bibitem [{\citenamefont {Sriram}\ \emph {et~al.}(2010)\citenamefont {Sriram},
  \citenamefont {Meyer},\ and\ \citenamefont {Furst}}]{Sriram:PoF_22:2010}%
  \BibitemOpen
  \bibfield  {author} {\bibinfo {author} {\bibfnamefont {I.}~\bibnamefont
  {Sriram}}, \bibinfo {author} {\bibfnamefont {A.}~\bibnamefont {Meyer}}, \
  and\ \bibinfo {author} {\bibfnamefont {E.~M.}\ \bibnamefont {Furst}},\ }\href
  {\doibase 10.1063/1.3450319} {\bibfield  {journal} {\bibinfo  {journal}
  {Phys. Fluids}\ }\textbf {\bibinfo {volume} {22}},\ \bibinfo {pages} {062003}
  (\bibinfo {year} {2010})}\BibitemShut {NoStop}%
\bibitem [{\citenamefont {Winter}\ \emph {et~al.}(2012)\citenamefont {Winter},
  \citenamefont {Horbach}, \citenamefont {Virnau},\ and\ \citenamefont
  {Binder}}]{Winter:PRL_108:2012}%
  \BibitemOpen
  \bibfield  {author} {\bibinfo {author} {\bibfnamefont {D.}~\bibnamefont
  {Winter}}, \bibinfo {author} {\bibfnamefont {J.}~\bibnamefont {Horbach}},
  \bibinfo {author} {\bibfnamefont {P.}~\bibnamefont {Virnau}}, \ and\ \bibinfo
  {author} {\bibfnamefont {K.}~\bibnamefont {Binder}},\ }\href {\doibase
  10.1103/PhysRevLett.108.028303} {\bibfield  {journal} {\bibinfo  {journal}
  {Phys. Rev. Lett.}\ }\textbf {\bibinfo {volume} {108}},\ \bibinfo {pages}
  {028303} (\bibinfo {year} {2012})}\BibitemShut {NoStop}%
\bibitem [{\citenamefont {Winter}\ and\ \citenamefont
  {Horbach}(2013)}]{Winter:JCP_138:2013}%
  \BibitemOpen
  \bibfield  {author} {\bibinfo {author} {\bibfnamefont {D.}~\bibnamefont
  {Winter}}\ and\ \bibinfo {author} {\bibfnamefont {J.}~\bibnamefont
  {Horbach}},\ }\href {\doibase 10.1063/1.4770335} {\bibfield  {journal}
  {\bibinfo  {journal} {J. Chem. Phys.}\ }\textbf {\bibinfo {volume} {138}},\
  \bibinfo {pages} {12A512} (\bibinfo {year} {2013})}\BibitemShut {NoStop}%
\bibitem [{\citenamefont {Horbach}\ \emph {et~al.}(2017)\citenamefont
  {Horbach}, \citenamefont {Siboni},\ and\ \citenamefont
  {Schnyder}}]{Horbach:EPJST_226:2017}%
  \BibitemOpen
  \bibfield  {author} {\bibinfo {author} {\bibfnamefont {J.}~\bibnamefont
  {Horbach}}, \bibinfo {author} {\bibfnamefont {N.~H.}\ \bibnamefont {Siboni}},
  \ and\ \bibinfo {author} {\bibfnamefont {S.~K.}\ \bibnamefont {Schnyder}},\
  }\href {\doibase 10.1140/epjst/e2017-70081-3} {\bibfield  {journal} {\bibinfo
   {journal} {The European Physical Journal Special Topics}\ }\textbf {\bibinfo
  {volume} {226}},\ \bibinfo {pages} {3113} (\bibinfo {year}
  {2017})}\BibitemShut {NoStop}%
\bibitem [{\citenamefont {Jack}\ \emph {et~al.}(2008)\citenamefont {Jack},
  \citenamefont {Kelsey}, \citenamefont {Garrahan},\ and\ \citenamefont
  {Chandler}}]{Jack:PRE_78:2008}%
  \BibitemOpen
  \bibfield  {author} {\bibinfo {author} {\bibfnamefont {R.~L.}\ \bibnamefont
  {Jack}}, \bibinfo {author} {\bibfnamefont {D.}~\bibnamefont {Kelsey}},
  \bibinfo {author} {\bibfnamefont {J.~P.}\ \bibnamefont {Garrahan}}, \ and\
  \bibinfo {author} {\bibfnamefont {D.}~\bibnamefont {Chandler}},\ }\href
  {\doibase 10.1103/PhysRevE.78.011506} {\bibfield  {journal} {\bibinfo
  {journal} {Phys. Rev. E}\ }\textbf {\bibinfo {volume} {78}},\ \bibinfo
  {pages} {011506} (\bibinfo {year} {2008})}\BibitemShut {NoStop}%
\bibitem [{\citenamefont {B\'enichou}\ \emph {et~al.}(2013)\citenamefont
  {B\'enichou}, \citenamefont {Bodrova}, \citenamefont {Chakraborty},
  \citenamefont {Illien}, \citenamefont {Law}, \citenamefont
  {Mej\'{\i}a-Monasterio}, \citenamefont {Oshanin},\ and\ \citenamefont
  {Voituriez}}]{Benichou:PRL_111:2013}%
  \BibitemOpen
  \bibfield  {author} {\bibinfo {author} {\bibfnamefont {O.}~\bibnamefont
  {B\'enichou}}, \bibinfo {author} {\bibfnamefont {A.}~\bibnamefont {Bodrova}},
  \bibinfo {author} {\bibfnamefont {D.}~\bibnamefont {Chakraborty}}, \bibinfo
  {author} {\bibfnamefont {P.}~\bibnamefont {Illien}}, \bibinfo {author}
  {\bibfnamefont {A.}~\bibnamefont {Law}}, \bibinfo {author} {\bibfnamefont
  {C.}~\bibnamefont {Mej\'{\i}a-Monasterio}}, \bibinfo {author} {\bibfnamefont
  {G.}~\bibnamefont {Oshanin}}, \ and\ \bibinfo {author} {\bibfnamefont
  {R.}~\bibnamefont {Voituriez}},\ }\href {\doibase
  10.1103/PhysRevLett.111.260601} {\bibfield  {journal} {\bibinfo  {journal}
  {Phys. Rev. Lett.}\ }\textbf {\bibinfo {volume} {111}},\ \bibinfo {pages}
  {260601} (\bibinfo {year} {2013})}\BibitemShut {NoStop}%
\bibitem [{\citenamefont {Illien}\ \emph {et~al.}(2013)\citenamefont {Illien},
  \citenamefont {B\'enichou}, \citenamefont {Mej\'{\i}a-Monasterio},
  \citenamefont {Oshanin},\ and\ \citenamefont
  {Voituriez}}]{Illien:PRL_111:2013}%
  \BibitemOpen
  \bibfield  {author} {\bibinfo {author} {\bibfnamefont {P.}~\bibnamefont
  {Illien}}, \bibinfo {author} {\bibfnamefont {O.}~\bibnamefont {B\'enichou}},
  \bibinfo {author} {\bibfnamefont {C.}~\bibnamefont {Mej\'{\i}a-Monasterio}},
  \bibinfo {author} {\bibfnamefont {G.}~\bibnamefont {Oshanin}}, \ and\
  \bibinfo {author} {\bibfnamefont {R.}~\bibnamefont {Voituriez}},\ }\href
  {\doibase 10.1103/PhysRevLett.111.038102} {\bibfield  {journal} {\bibinfo
  {journal} {Phys. Rev. Lett.}\ }\textbf {\bibinfo {volume} {111}},\ \bibinfo
  {pages} {038102} (\bibinfo {year} {2013})}\BibitemShut {NoStop}%
\bibitem [{\citenamefont {Leitmann}\ and\ \citenamefont
  {Franosch}(2013)}]{Leitmann:PRL_111:2013}%
  \BibitemOpen
  \bibfield  {author} {\bibinfo {author} {\bibfnamefont {S.}~\bibnamefont
  {Leitmann}}\ and\ \bibinfo {author} {\bibfnamefont {T.}~\bibnamefont
  {Franosch}},\ }\href {\doibase 10.1103/PhysRevLett.111.190603} {\bibfield
  {journal} {\bibinfo  {journal} {Phys. Rev. Lett.}\ }\textbf {\bibinfo
  {volume} {111}},\ \bibinfo {pages} {190603} (\bibinfo {year}
  {2013})}\BibitemShut {NoStop}%
\bibitem [{\citenamefont {Basu}\ and\ \citenamefont
  {Maes}(2014)}]{Basu:JPAMT_47:2014}%
  \BibitemOpen
  \bibfield  {author} {\bibinfo {author} {\bibfnamefont {U.}~\bibnamefont
  {Basu}}\ and\ \bibinfo {author} {\bibfnamefont {C.}~\bibnamefont {Maes}},\
  }\href {\doibase 10.1088/1751-8113/47/25/255003} {\bibfield  {journal}
  {\bibinfo  {journal} {J. Phys. A}\ }\textbf {\bibinfo {volume} {47}},\
  \bibinfo {pages} {255003} (\bibinfo {year} {2014})}\BibitemShut {NoStop}%
\bibitem [{\citenamefont {Illien}\ \emph {et~al.}(2014)\citenamefont {Illien},
  \citenamefont {B\'enichou}, \citenamefont {Oshanin},\ and\ \citenamefont
  {Voituriez}}]{Illien:PRL_113:2014}%
  \BibitemOpen
  \bibfield  {author} {\bibinfo {author} {\bibfnamefont {P.}~\bibnamefont
  {Illien}}, \bibinfo {author} {\bibfnamefont {O.}~\bibnamefont {B\'enichou}},
  \bibinfo {author} {\bibfnamefont {G.}~\bibnamefont {Oshanin}}, \ and\
  \bibinfo {author} {\bibfnamefont {R.}~\bibnamefont {Voituriez}},\ }\href
  {\doibase 10.1103/PhysRevLett.113.030603} {\bibfield  {journal} {\bibinfo
  {journal} {Phys. Rev. Lett.}\ }\textbf {\bibinfo {volume} {113}},\ \bibinfo
  {pages} {030603} (\bibinfo {year} {2014})}\BibitemShut {NoStop}%
\bibitem [{\citenamefont {B\'enichou}\ \emph {et~al.}(2014)\citenamefont
  {B\'enichou}, \citenamefont {Illien}, \citenamefont {Oshanin}, \citenamefont
  {Sarracino},\ and\ \citenamefont {Voituriez}}]{Benichou:PRL_113:2014}%
  \BibitemOpen
  \bibfield  {author} {\bibinfo {author} {\bibfnamefont {O.}~\bibnamefont
  {B\'enichou}}, \bibinfo {author} {\bibfnamefont {P.}~\bibnamefont {Illien}},
  \bibinfo {author} {\bibfnamefont {G.}~\bibnamefont {Oshanin}}, \bibinfo
  {author} {\bibfnamefont {A.}~\bibnamefont {Sarracino}}, \ and\ \bibinfo
  {author} {\bibfnamefont {R.}~\bibnamefont {Voituriez}},\ }\href {\doibase
  10.1103/PhysRevLett.113.268002} {\bibfield  {journal} {\bibinfo  {journal}
  {Phys. Rev. Lett.}\ }\textbf {\bibinfo {volume} {113}},\ \bibinfo {pages}
  {268002} (\bibinfo {year} {2014})}\BibitemShut {NoStop}%
\bibitem [{\citenamefont {Illien}\ \emph {et~al.}()\citenamefont {Illien},
  \citenamefont {Bénichou}, \citenamefont {Oshanin},\ and\ \citenamefont
  {Voituriez}}]{Illien:JSMTE:2015}%
  \BibitemOpen
  \bibfield  {author} {\bibinfo {author} {\bibfnamefont {P.}~\bibnamefont
  {Illien}}, \bibinfo {author} {\bibfnamefont {O.}~\bibnamefont {Bénichou}},
  \bibinfo {author} {\bibfnamefont {G.}~\bibnamefont {Oshanin}}, \ and\
  \bibinfo {author} {\bibfnamefont {R.}~\bibnamefont {Voituriez}},\ }\href
  {\doibase 10.1088/1742-5468/2015/11/P11016} {\bibfield  {journal} {\bibinfo
  {journal} {J. Stat. Mech.}\ }\textbf {\bibinfo {volume} {(2015)}},\ \bibinfo
  {pages} {P11016}}\BibitemShut {NoStop}%
\bibitem [{\citenamefont {Baiesi}\ \emph {et~al.}(2015)\citenamefont {Baiesi},
  \citenamefont {Stella},\ and\ \citenamefont
  {Vanderzande}}]{Baiesi:PRE_92:2015}%
  \BibitemOpen
  \bibfield  {author} {\bibinfo {author} {\bibfnamefont {M.}~\bibnamefont
  {Baiesi}}, \bibinfo {author} {\bibfnamefont {A.~L.}\ \bibnamefont {Stella}},
  \ and\ \bibinfo {author} {\bibfnamefont {C.}~\bibnamefont {Vanderzande}},\
  }\href {\doibase 10.1103/PhysRevE.92.042121} {\bibfield  {journal} {\bibinfo
  {journal} {Phys. Rev. E}\ }\textbf {\bibinfo {volume} {92}},\ \bibinfo
  {pages} {042121} (\bibinfo {year} {2015})}\BibitemShut {NoStop}%
\bibitem [{\citenamefont {B\'enichou}\ \emph {et~al.}(2016)\citenamefont
  {B\'enichou}, \citenamefont {Illien}, \citenamefont {Oshanin}, \citenamefont
  {Sarracino},\ and\ \citenamefont {Voituriez}}]{Benichou:PRE_93:2016}%
  \BibitemOpen
  \bibfield  {author} {\bibinfo {author} {\bibfnamefont {O.}~\bibnamefont
  {B\'enichou}}, \bibinfo {author} {\bibfnamefont {P.}~\bibnamefont {Illien}},
  \bibinfo {author} {\bibfnamefont {G.}~\bibnamefont {Oshanin}}, \bibinfo
  {author} {\bibfnamefont {A.}~\bibnamefont {Sarracino}}, \ and\ \bibinfo
  {author} {\bibfnamefont {R.}~\bibnamefont {Voituriez}},\ }\href {\doibase
  10.1103/PhysRevE.93.032128} {\bibfield  {journal} {\bibinfo  {journal} {Phys.
  Rev. E}\ }\textbf {\bibinfo {volume} {93}},\ \bibinfo {pages} {032128}
  (\bibinfo {year} {2016})}\BibitemShut {NoStop}%
\bibitem [{\citenamefont {Leitmann}\ and\ \citenamefont
  {Franosch}(2017)}]{Leitmann:PRL_118:2017}%
  \BibitemOpen
  \bibfield  {author} {\bibinfo {author} {\bibfnamefont {S.}~\bibnamefont
  {Leitmann}}\ and\ \bibinfo {author} {\bibfnamefont {T.}~\bibnamefont
  {Franosch}},\ }\href {\doibase 10.1103/PhysRevLett.118.018001} {\bibfield
  {journal} {\bibinfo  {journal} {Phys. Rev. Lett.}\ }\textbf {\bibinfo
  {volume} {118}},\ \bibinfo {pages} {018001} (\bibinfo {year}
  {2017})}\BibitemShut {NoStop}%
\bibitem [{\citenamefont {Schroer}\ and\ \citenamefont
  {Heuer}(2013{\natexlab{a}})}]{Schroer:PRL_110:2013}%
  \BibitemOpen
  \bibfield  {author} {\bibinfo {author} {\bibfnamefont {C.~F.~E.}\
  \bibnamefont {Schroer}}\ and\ \bibinfo {author} {\bibfnamefont
  {A.}~\bibnamefont {Heuer}},\ }\href {\doibase 10.1103/PhysRevLett.110.067801}
  {\bibfield  {journal} {\bibinfo  {journal} {Phys. Rev. Lett.}\ }\textbf
  {\bibinfo {volume} {110}},\ \bibinfo {pages} {067801} (\bibinfo {year}
  {2013}{\natexlab{a}})}\BibitemShut {NoStop}%
\bibitem [{\citenamefont {Schroer}\ and\ \citenamefont
  {Heuer}(2013{\natexlab{b}})}]{Schroer:JCP_138:2013}%
  \BibitemOpen
  \bibfield  {author} {\bibinfo {author} {\bibfnamefont {C.~F.~E.}\
  \bibnamefont {Schroer}}\ and\ \bibinfo {author} {\bibfnamefont
  {A.}~\bibnamefont {Heuer}},\ }\href {\doibase 10.1063/1.4772627} {\bibfield
  {journal} {\bibinfo  {journal} {J. Chem. Phys.}\ }\textbf {\bibinfo {volume}
  {138}},\ \bibinfo {pages} {12A518} (\bibinfo {year}
  {2013}{\natexlab{b}})}\BibitemShut {NoStop}%
\bibitem [{\citenamefont {Burioni}\ \emph {et~al.}(2014)\citenamefont
  {Burioni}, \citenamefont {Gradenigo}, \citenamefont {Sarracino},
  \citenamefont {Vezzani},\ and\ \citenamefont
  {Vulpiani}}]{Burioni:CTP_62:2014}%
  \BibitemOpen
  \bibfield  {author} {\bibinfo {author} {\bibfnamefont {R.}~\bibnamefont
  {Burioni}}, \bibinfo {author} {\bibfnamefont {G.}~\bibnamefont {Gradenigo}},
  \bibinfo {author} {\bibfnamefont {A.}~\bibnamefont {Sarracino}}, \bibinfo
  {author} {\bibfnamefont {A.}~\bibnamefont {Vezzani}}, \ and\ \bibinfo
  {author} {\bibfnamefont {A.}~\bibnamefont {Vulpiani}},\ }\href {\doibase
  10.1088/0253-6102/62/4/09} {\bibfield  {journal} {\bibinfo  {journal}
  {Commun. Theor. Phys.}\ }\textbf {\bibinfo {volume} {62}},\ \bibinfo {pages}
  {514} (\bibinfo {year} {2014})}\BibitemShut {NoStop}%
\bibitem [{\citenamefont {Démery}\ \emph {et~al.}(2014)\citenamefont
  {Démery}, \citenamefont {Bénichou},\ and\ \citenamefont
  {Jacquin}}]{Demery:NJP_16:2014}%
  \BibitemOpen
  \bibfield  {author} {\bibinfo {author} {\bibfnamefont {V.}~\bibnamefont
  {Démery}}, \bibinfo {author} {\bibfnamefont {O.}~\bibnamefont {Bénichou}},
  \ and\ \bibinfo {author} {\bibfnamefont {H.}~\bibnamefont {Jacquin}},\ }\href
  {\doibase 10.1088/1367-2630/16/5/053032} {\bibfield  {journal} {\bibinfo
  {journal} {New J. Phys.}\ }\textbf {\bibinfo {volume} {16}},\ \bibinfo
  {pages} {053032} (\bibinfo {year} {2014})}\BibitemShut {NoStop}%
\bibitem [{\citenamefont {D\'emery}(2015)}]{Demery:PRE_91:2015}%
  \BibitemOpen
  \bibfield  {author} {\bibinfo {author} {\bibfnamefont {V.}~\bibnamefont
  {D\'emery}},\ }\href {\doibase 10.1103/PhysRevE.91.062301} {\bibfield
  {journal} {\bibinfo  {journal} {Phys. Rev. E}\ }\textbf {\bibinfo {volume}
  {91}},\ \bibinfo {pages} {062301} (\bibinfo {year} {2015})}\BibitemShut
  {NoStop}%
\bibitem [{\citenamefont {Wang}\ and\ \citenamefont
  {Sperl}(2016)}]{Wang:PRE_93:2016}%
  \BibitemOpen
  \bibfield  {author} {\bibinfo {author} {\bibfnamefont {T.}~\bibnamefont
  {Wang}}\ and\ \bibinfo {author} {\bibfnamefont {M.}~\bibnamefont {Sperl}},\
  }\href {\doibase 10.1103/PhysRevE.93.022606} {\bibfield  {journal} {\bibinfo
  {journal} {Phys. Rev. E}\ }\textbf {\bibinfo {volume} {93}},\ \bibinfo
  {pages} {022606} (\bibinfo {year} {2016})}\BibitemShut {NoStop}%
\bibitem [{\citenamefont {Gazuz}\ \emph {et~al.}(2009)\citenamefont {Gazuz},
  \citenamefont {Puertas}, \citenamefont {Voigtmann},\ and\ \citenamefont
  {Fuchs}}]{Gazuz:PRL_102:2009}%
  \BibitemOpen
  \bibfield  {author} {\bibinfo {author} {\bibfnamefont {I.}~\bibnamefont
  {Gazuz}}, \bibinfo {author} {\bibfnamefont {A.~M.}\ \bibnamefont {Puertas}},
  \bibinfo {author} {\bibfnamefont {{\relax Th}.}~\bibnamefont {Voigtmann}}, \
  and\ \bibinfo {author} {\bibfnamefont {M.}~\bibnamefont {Fuchs}},\ }\href
  {\doibase 10.1103/PhysRevLett.102.248302} {\bibfield  {journal} {\bibinfo
  {journal} {Phys. Rev. Lett.}\ }\textbf {\bibinfo {volume} {102}},\ \bibinfo
  {pages} {248302} (\bibinfo {year} {2009})}\BibitemShut {NoStop}%
\bibitem [{\citenamefont {Gnann}\ \emph {et~al.}(2011)\citenamefont {Gnann},
  \citenamefont {Gazuz}, \citenamefont {Puertas}, \citenamefont {Fuchs},\ and\
  \citenamefont {Voigtmann}}]{Gnann:SM_7:2011}%
  \BibitemOpen
  \bibfield  {author} {\bibinfo {author} {\bibfnamefont {M.~V.}\ \bibnamefont
  {Gnann}}, \bibinfo {author} {\bibfnamefont {I.}~\bibnamefont {Gazuz}},
  \bibinfo {author} {\bibfnamefont {A.~M.}\ \bibnamefont {Puertas}}, \bibinfo
  {author} {\bibfnamefont {M.}~\bibnamefont {Fuchs}}, \ and\ \bibinfo {author}
  {\bibfnamefont {{\relax Th}.}~\bibnamefont {Voigtmann}},\ }\href {\doibase
  10.1039/C0SM00828A} {\bibfield  {journal} {\bibinfo  {journal} {Soft Matter}\
  }\textbf {\bibinfo {volume} {7}},\ \bibinfo {pages} {1390} (\bibinfo {year}
  {2011})}\BibitemShut {NoStop}%
\bibitem [{\citenamefont {Gnann}\ and\ \citenamefont
  {Voigtmann}(2012)}]{Gnann:PRE_86:2012}%
  \BibitemOpen
  \bibfield  {author} {\bibinfo {author} {\bibfnamefont {M.~V.}\ \bibnamefont
  {Gnann}}\ and\ \bibinfo {author} {\bibfnamefont {{\relax Th}.}~\bibnamefont
  {Voigtmann}},\ }\href {\doibase 10.1103/PhysRevE.86.011406} {\bibfield
  {journal} {\bibinfo  {journal} {Phys. Rev. E}\ }\textbf {\bibinfo {volume}
  {86}},\ \bibinfo {pages} {011406} (\bibinfo {year} {2012})}\BibitemShut
  {NoStop}%
\bibitem [{\citenamefont {Harrer}\ \emph {et~al.}(2012)\citenamefont {Harrer},
  \citenamefont {Winter}, \citenamefont {Horbach}, \citenamefont {Fuchs},\ and\
  \citenamefont {Voigtmann}}]{Harrer:JPCM_24:2012}%
  \BibitemOpen
  \bibfield  {author} {\bibinfo {author} {\bibfnamefont {C.~J.}\ \bibnamefont
  {Harrer}}, \bibinfo {author} {\bibfnamefont {D.}~\bibnamefont {Winter}},
  \bibinfo {author} {\bibfnamefont {J.}~\bibnamefont {Horbach}}, \bibinfo
  {author} {\bibfnamefont {M.}~\bibnamefont {Fuchs}}, \ and\ \bibinfo {author}
  {\bibfnamefont {{\relax Th}.}~\bibnamefont {Voigtmann}},\ }\href {\doibase
  10.1088/0953-8984/24/46/464105} {\bibfield  {journal} {\bibinfo  {journal}
  {J. Phys. Condens. Matter}\ }\textbf {\bibinfo {volume} {24}},\ \bibinfo
  {pages} {464105} (\bibinfo {year} {2012})}\BibitemShut {NoStop}%
\bibitem [{\citenamefont {Gazuz}\ and\ \citenamefont
  {Fuchs}(2013)}]{Gazuz:PRE_87:2013}%
  \BibitemOpen
  \bibfield  {author} {\bibinfo {author} {\bibfnamefont {I.}~\bibnamefont
  {Gazuz}}\ and\ \bibinfo {author} {\bibfnamefont {M.}~\bibnamefont {Fuchs}},\
  }\href {\doibase 10.1103/PhysRevE.87.032304} {\bibfield  {journal} {\bibinfo
  {journal} {Phys. Rev. E}\ }\textbf {\bibinfo {volume} {87}},\ \bibinfo
  {pages} {032304} (\bibinfo {year} {2013})}\BibitemShut {NoStop}%
\bibitem [{\citenamefont {Wang}\ \emph {et~al.}(2014)\citenamefont {Wang},
  \citenamefont {Grob}, \citenamefont {Zippelius},\ and\ \citenamefont
  {Sperl}}]{Wang:PRE_89:2014}%
  \BibitemOpen
  \bibfield  {author} {\bibinfo {author} {\bibfnamefont {T.}~\bibnamefont
  {Wang}}, \bibinfo {author} {\bibfnamefont {M.}~\bibnamefont {Grob}}, \bibinfo
  {author} {\bibfnamefont {A.}~\bibnamefont {Zippelius}}, \ and\ \bibinfo
  {author} {\bibfnamefont {M.}~\bibnamefont {Sperl}},\ }\href {\doibase
  10.1103/PhysRevE.89.042209} {\bibfield  {journal} {\bibinfo  {journal} {Phys.
  Rev. E}\ }\textbf {\bibinfo {volume} {89}},\ \bibinfo {pages} {042209}
  (\bibinfo {year} {2014})}\BibitemShut {NoStop}%
\bibitem [{\citenamefont {Gruber}\ \emph {et~al.}(2016)\citenamefont {Gruber},
  \citenamefont {Abade}, \citenamefont {Puertas},\ and\ \citenamefont
  {Fuchs}}]{Gruber:PRE_94:2016}%
  \BibitemOpen
  \bibfield  {author} {\bibinfo {author} {\bibfnamefont {M.}~\bibnamefont
  {Gruber}}, \bibinfo {author} {\bibfnamefont {G.~C.}\ \bibnamefont {Abade}},
  \bibinfo {author} {\bibfnamefont {A.~M.}\ \bibnamefont {Puertas}}, \ and\
  \bibinfo {author} {\bibfnamefont {M.}~\bibnamefont {Fuchs}},\ }\href
  {\doibase 10.1103/PhysRevE.94.042602} {\bibfield  {journal} {\bibinfo
  {journal} {Phys. Rev. E}\ }\textbf {\bibinfo {volume} {94}},\ \bibinfo
  {pages} {042602} (\bibinfo {year} {2016})}\BibitemShut {NoStop}%
\bibitem [{\citenamefont {Squires}\ and\ \citenamefont
  {Brady}(2005)}]{Squires:PoF_17:2005}%
  \BibitemOpen
  \bibfield  {author} {\bibinfo {author} {\bibfnamefont {T.~M.}\ \bibnamefont
  {Squires}}\ and\ \bibinfo {author} {\bibfnamefont {J.~F.}\ \bibnamefont
  {Brady}},\ }\href {\doibase 10.1063/1.1960607} {\bibfield  {journal}
  {\bibinfo  {journal} {Phys. Fluids}\ }\textbf {\bibinfo {volume} {17}},\
  \bibinfo {pages} {073101} (\bibinfo {year} {2005})}\BibitemShut {NoStop}%
\bibitem [{\citenamefont {Khair}\ and\ \citenamefont
  {Brady}(2006)}]{Khair:JFM_557:2006}%
  \BibitemOpen
  \bibfield  {author} {\bibinfo {author} {\bibfnamefont {A.~S.}\ \bibnamefont
  {Khair}}\ and\ \bibinfo {author} {\bibfnamefont {J.~F.}\ \bibnamefont
  {Brady}},\ }\href {\doibase 10.1017/S0022112006009608} {\bibfield  {journal}
  {\bibinfo  {journal} {J. Fluid Mech.}\ }\textbf {\bibinfo {volume} {557}},\
  \bibinfo {pages} {73} (\bibinfo {year} {2006})}\BibitemShut {NoStop}%
\bibitem [{\citenamefont {Zia}\ and\ \citenamefont
  {Brady}(2010)}]{Zia:JFM_658:2010}%
  \BibitemOpen
  \bibfield  {author} {\bibinfo {author} {\bibfnamefont {R.~N.}\ \bibnamefont
  {Zia}}\ and\ \bibinfo {author} {\bibfnamefont {J.~F.}\ \bibnamefont
  {Brady}},\ }\href {\doibase 10.1017/S0022112010001606} {\bibfield  {journal}
  {\bibinfo  {journal} {J. Fluid Mech.}\ }\textbf {\bibinfo {volume} {658}},\
  \bibinfo {pages} {188} (\bibinfo {year} {2010})}\BibitemShut {NoStop}%
\bibitem [{\citenamefont {Swan}\ and\ \citenamefont
  {Zia}(2013)}]{Swan:PoF_25:2013}%
  \BibitemOpen
  \bibfield  {author} {\bibinfo {author} {\bibfnamefont {J.~W.}\ \bibnamefont
  {Swan}}\ and\ \bibinfo {author} {\bibfnamefont {R.~N.}\ \bibnamefont {Zia}},\
  }\href {\doibase 10.1063/1.4818810} {\bibfield  {journal} {\bibinfo
  {journal} {Phys. Fluids}\ }\textbf {\bibinfo {volume} {25}},\ \bibinfo
  {pages} {083303} (\bibinfo {year} {2013})}\BibitemShut {NoStop}%
\bibitem [{\citenamefont {Hoh}\ and\ \citenamefont
  {Zia}(2016)}]{Hoh:JFM_795:2016}%
  \BibitemOpen
  \bibfield  {author} {\bibinfo {author} {\bibfnamefont {N.~J.}\ \bibnamefont
  {Hoh}}\ and\ \bibinfo {author} {\bibfnamefont {R.~N.}\ \bibnamefont {Zia}},\
  }\href {\doibase 10.1017/jfm.2016.209} {\bibfield  {journal} {\bibinfo
  {journal} {J. Fluid Mech.}\ }\textbf {\bibinfo {volume} {795}},\ \bibinfo
  {pages} {739} (\bibinfo {year} {2016})}\BibitemShut {NoStop}%
\bibitem [{\citenamefont {Ballentine}(2003)}]{Ballentine:WS:2003}%
  \BibitemOpen
  \bibfield  {author} {\bibinfo {author} {\bibfnamefont {L.~E.}\ \bibnamefont
  {Ballentine}},\ }\href@noop {} {\emph {\bibinfo {title} {Quantum Mechanics: A
  Modern Development}}}\ (\bibinfo  {publisher} {World Scientific},\ \bibinfo
  {address} {Singapore},\ \bibinfo {year} {2003})\BibitemShut {NoStop}%
\bibitem [{\citenamefont {van Rossum}\ and\ \citenamefont
  {Nieuwenhuizen}(1999)}]{Rossum:RMP_71:1999}%
  \BibitemOpen
  \bibfield  {author} {\bibinfo {author} {\bibfnamefont {M.~C.~W.}\
  \bibnamefont {van Rossum}}\ and\ \bibinfo {author} {\bibfnamefont {{\relax
  Th}.~M.}\ \bibnamefont {Nieuwenhuizen}},\ }\href {\doibase
  10.1103/RevModPhys.71.313} {\bibfield  {journal} {\bibinfo  {journal} {Rev.
  Mod. Phys.}\ }\textbf {\bibinfo {volume} {71}},\ \bibinfo {pages} {313}
  (\bibinfo {year} {1999})}\BibitemShut {NoStop}%
\bibitem [{\citenamefont {Montroll}\ and\ \citenamefont
  {Scher}(1973)}]{Montroll:JSP_9:1973}%
  \BibitemOpen
  \bibfield  {author} {\bibinfo {author} {\bibfnamefont {E.~W.}\ \bibnamefont
  {Montroll}}\ and\ \bibinfo {author} {\bibfnamefont {H.}~\bibnamefont
  {Scher}},\ }\href {\doibase 10.1007/BF01016843} {\bibfield  {journal}
  {\bibinfo  {journal} {Journal of Statistical Physics}\ }\textbf {\bibinfo
  {volume} {9}},\ \bibinfo {pages} {101} (\bibinfo {year} {1973})}\BibitemShut
  {NoStop}%
\bibitem [{\citenamefont {Haus}\ and\ \citenamefont
  {Kehr}(1987)}]{Haus:PR_150:1987}%
  \BibitemOpen
  \bibfield  {author} {\bibinfo {author} {\bibfnamefont {J.}~\bibnamefont
  {Haus}}\ and\ \bibinfo {author} {\bibfnamefont {K.}~\bibnamefont {Kehr}},\
  }\href {\doibase 10.1016/0370-1573(87)90005-6} {\bibfield  {journal}
  {\bibinfo  {journal} {Phys. Rep.}\ }\textbf {\bibinfo {volume} {150}},\
  \bibinfo {pages} {263 } (\bibinfo {year} {1987})}\BibitemShut {NoStop}%
\bibitem [{\citenamefont {Nieuwenhuizen}\ \emph {et~al.}(1986)\citenamefont
  {Nieuwenhuizen}, \citenamefont {van Velthoven},\ and\ \citenamefont
  {Ernst}}]{Nieuwenhuizen:PRL_57:1986}%
  \BibitemOpen
  \bibfield  {author} {\bibinfo {author} {\bibfnamefont {{\relax Th}.~M.}\
  \bibnamefont {Nieuwenhuizen}}, \bibinfo {author} {\bibfnamefont {P.~F.~J.}\
  \bibnamefont {van Velthoven}}, \ and\ \bibinfo {author} {\bibfnamefont
  {M.~H.}\ \bibnamefont {Ernst}},\ }\href {\doibase
  10.1103/PhysRevLett.57.2477} {\bibfield  {journal} {\bibinfo  {journal}
  {Phys. Rev. Lett.}\ }\textbf {\bibinfo {volume} {57}},\ \bibinfo {pages}
  {2477} (\bibinfo {year} {1986})}\BibitemShut {NoStop}%
\bibitem [{\citenamefont {Nieuwenhuizen}\ \emph {et~al.}(1987)\citenamefont
  {Nieuwenhuizen}, \citenamefont {van Velthoven},\ and\ \citenamefont
  {Ernst}}]{Nieuwenhuizen:JPAMG_20:1987}%
  \BibitemOpen
  \bibfield  {author} {\bibinfo {author} {\bibfnamefont {{\relax Th}.~M.}\
  \bibnamefont {Nieuwenhuizen}}, \bibinfo {author} {\bibfnamefont {P.~F.~J.}\
  \bibnamefont {van Velthoven}}, \ and\ \bibinfo {author} {\bibfnamefont
  {M.~H.}\ \bibnamefont {Ernst}},\ }\href {\doibase
  10.1088/0305-4470/20/12/044} {\bibfield  {journal} {\bibinfo  {journal} {J.
  Phys. A}\ }\textbf {\bibinfo {volume} {20}},\ \bibinfo {pages} {4001}
  (\bibinfo {year} {1987})}\BibitemShut {NoStop}%
\bibitem [{\citenamefont {Ernst}\ \emph {et~al.}(1987)\citenamefont {Ernst},
  \citenamefont {Nieuwenhuizen},\ and\ \citenamefont {van
  Velthoven}}]{Ernst:JPAMG_20:1987}%
  \BibitemOpen
  \bibfield  {author} {\bibinfo {author} {\bibfnamefont {M.~H.}\ \bibnamefont
  {Ernst}}, \bibinfo {author} {\bibfnamefont {{\relax Th}.~M.}\ \bibnamefont
  {Nieuwenhuizen}}, \ and\ \bibinfo {author} {\bibfnamefont {P.~F.~J.}\
  \bibnamefont {van Velthoven}},\ }\href {\doibase 10.1088/0305-4470/20/15/045}
  {\bibfield  {journal} {\bibinfo  {journal} {J. Phys. A}\ }\textbf {\bibinfo
  {volume} {20}},\ \bibinfo {pages} {5335} (\bibinfo {year}
  {1987})}\BibitemShut {NoStop}%
\bibitem [{\citenamefont {Nieuwenhuizen}(1989)}]{Nieuwenhuizen:PA_157:1989}%
  \BibitemOpen
  \bibfield  {author} {\bibinfo {author} {\bibfnamefont {{\relax Th}.~M.}\
  \bibnamefont {Nieuwenhuizen}},\ }\href {\doibase
  https://doi.org/10.1016/0378-4371(89)90036-8} {\bibfield  {journal} {\bibinfo
   {journal} {Physica A}\ }\textbf {\bibinfo {volume} {157}},\ \bibinfo {pages}
  {1101 } (\bibinfo {year} {1989})}\BibitemShut {NoStop}%
\bibitem [{\citenamefont {Wortis}(1963)}]{Wortis:PRE_132:1963}%
  \BibitemOpen
  \bibfield  {author} {\bibinfo {author} {\bibfnamefont {M.}~\bibnamefont
  {Wortis}},\ }\href {\doibase 10.1103/PhysRev.132.85} {\bibfield  {journal}
  {\bibinfo  {journal} {Phys. Rev.}\ }\textbf {\bibinfo {volume} {132}},\
  \bibinfo {pages} {85} (\bibinfo {year} {1963})}\BibitemShut {NoStop}%
\bibitem [{\citenamefont {Gradshteyn}\ and\ \citenamefont
  {Ryzhik}(2007)}]{Gradshteyn:AP:2007}%
  \BibitemOpen
  \bibfield  {author} {\bibinfo {author} {\bibfnamefont {I.}~\bibnamefont
  {Gradshteyn}}\ and\ \bibinfo {author} {\bibfnamefont {I.}~\bibnamefont
  {Ryzhik}},\ }\href@noop {} {\emph {\bibinfo {title} {Table of integrals,
  series and products}}},\ \bibinfo {edition} {7th}\ ed.,\ edited by\ \bibinfo
  {editor} {\bibfnamefont {A.}~\bibnamefont {Jeffrey}}\ and\ \bibinfo {editor}
  {\bibfnamefont {D.}~\bibnamefont {Zwillinger}}\ (\bibinfo  {publisher}
  {Academic Press},\ \bibinfo {address} {Amsterdam},\ \bibinfo {year}
  {2007})\BibitemShut {NoStop}%
\bibitem [{\citenamefont {Ernst}\ and\ \citenamefont {van
  Beijeren}(1981)}]{Ernst:JSP_26:1981}%
  \BibitemOpen
  \bibfield  {author} {\bibinfo {author} {\bibfnamefont {M.~H.}\ \bibnamefont
  {Ernst}}\ and\ \bibinfo {author} {\bibfnamefont {H.}~\bibnamefont {van
  Beijeren}},\ }\href {\doibase 10.1007/BF01106782} {\bibfield  {journal}
  {\bibinfo  {journal} {Journal of Statistical Physics}\ }\textbf {\bibinfo
  {volume} {26}},\ \bibinfo {pages} {1} (\bibinfo {year} {1981})}\BibitemShut
  {NoStop}%
\bibitem [{\citenamefont {van Beijeren}(1982)}]{vanBeijeren:RMP_54:1982}%
  \BibitemOpen
  \bibfield  {author} {\bibinfo {author} {\bibfnamefont {H.}~\bibnamefont {van
  Beijeren}},\ }\href {\doibase 10.1103/RevModPhys.54.195} {\bibfield
  {journal} {\bibinfo  {journal} {Rev. Mod. Phys.}\ }\textbf {\bibinfo {volume}
  {54}},\ \bibinfo {pages} {195} (\bibinfo {year} {1982})}\BibitemShut
  {NoStop}%
\bibitem [{\citenamefont {Joyce}(2001)}]{Joyce:JPAMG_34:2001}%
  \BibitemOpen
  \bibfield  {author} {\bibinfo {author} {\bibfnamefont {G.~S.}\ \bibnamefont
  {Joyce}},\ }\href {\doibase 10.1088/0305-4470/34/18/311} {\bibfield
  {journal} {\bibinfo  {journal} {J. Phys. A}\ }\textbf {\bibinfo {volume}
  {34}},\ \bibinfo {pages} {3831} (\bibinfo {year} {2001})}\BibitemShut
  {NoStop}%
\bibitem [{\citenamefont {Joyce}(2002)}]{Joyce:JPAMG_35:2002}%
  \BibitemOpen
  \bibfield  {author} {\bibinfo {author} {\bibfnamefont {G.~S.}\ \bibnamefont
  {Joyce}},\ }\href {\doibase 10.1088/0305-4470/35/46/307} {\bibfield
  {journal} {\bibinfo  {journal} {J. Phys. A}\ }\textbf {\bibinfo {volume}
  {35}},\ \bibinfo {pages} {9811} (\bibinfo {year} {2002})}\BibitemShut
  {NoStop}%
\bibitem [{\citenamefont {Stauffer}\ and\ \citenamefont
  {Aharony}(1994)}]{Stauffer:Taylor:1994}%
  \BibitemOpen
  \bibfield  {author} {\bibinfo {author} {\bibfnamefont {D.}~\bibnamefont
  {Stauffer}}\ and\ \bibinfo {author} {\bibfnamefont {A.}~\bibnamefont
  {Aharony}},\ }\href@noop {} {\emph {\bibinfo {title} {Introduction to
  Percolation Theory}}}\ (\bibinfo  {publisher} {Taylor and Francis},\ \bibinfo
  {address} {London},\ \bibinfo {year} {1994})\BibitemShut {NoStop}%
\bibitem [{\citenamefont {ben Avraham}\ and\ \citenamefont
  {Havlin}(2000)}]{benAvraham:CUP:2000}%
  \BibitemOpen
  \bibfield  {author} {\bibinfo {author} {\bibfnamefont {D.}~\bibnamefont {ben
  Avraham}}\ and\ \bibinfo {author} {\bibfnamefont {S.}~\bibnamefont
  {Havlin}},\ }\href@noop {} {\emph {\bibinfo {title} {Diffusion and Reactions
  in Fractals and Disordered Systems}}}\ (\bibinfo  {publisher} {Cambridge
  University Press},\ \bibinfo {address} {Cambridge},\ \bibinfo {year}
  {2000})\BibitemShut {NoStop}%
\bibitem [{\citenamefont {Kammerer}\ \emph {et~al.}(2008)\citenamefont
  {Kammerer}, \citenamefont {Höfling},\ and\ \citenamefont
  {Franosch}}]{Kammerer:EPL_84:2008}%
  \BibitemOpen
  \bibfield  {author} {\bibinfo {author} {\bibfnamefont {A.}~\bibnamefont
  {Kammerer}}, \bibinfo {author} {\bibfnamefont {F.}~\bibnamefont {Höfling}},
  \ and\ \bibinfo {author} {\bibfnamefont {T.}~\bibnamefont {Franosch}},\
  }\href {\doibase 10.1209/0295-5075/84/66002} {\bibfield  {journal} {\bibinfo
  {journal} {EPL (Europhysics Letters)}\ }\textbf {\bibinfo {volume} {84}},\
  \bibinfo {pages} {66002} (\bibinfo {year} {2008})}\BibitemShut {NoStop}%
\bibitem [{\citenamefont {van Beijeren}\ \emph {et~al.}(1985)\citenamefont {van
  Beijeren}, \citenamefont {Kutner},\ and\ \citenamefont
  {Spohn}}]{vanBeijeren:PRL_54:1985}%
  \BibitemOpen
  \bibfield  {author} {\bibinfo {author} {\bibfnamefont {H.}~\bibnamefont {van
  Beijeren}}, \bibinfo {author} {\bibfnamefont {R.}~\bibnamefont {Kutner}}, \
  and\ \bibinfo {author} {\bibfnamefont {H.}~\bibnamefont {Spohn}},\ }\href
  {\doibase 10.1103/PhysRevLett.54.2026} {\bibfield  {journal} {\bibinfo
  {journal} {Phys. Rev. Lett.}\ }\textbf {\bibinfo {volume} {54}},\ \bibinfo
  {pages} {2026} (\bibinfo {year} {1985})}\BibitemShut {NoStop}%
\bibitem [{\citenamefont {Yau}(2004)}]{Yau:AoM_159:2004}%
  \BibitemOpen
  \bibfield  {author} {\bibinfo {author} {\bibfnamefont {H.-T.}\ \bibnamefont
  {Yau}},\ }\href {http://www.jstor.org/stable/3597254} {\bibfield  {journal}
  {\bibinfo  {journal} {Annals of Mathematics}\ }\textbf {\bibinfo {volume}
  {159}},\ \bibinfo {pages} {377} (\bibinfo {year} {2004})}\BibitemShut
  {NoStop}%
\bibitem [{\citenamefont {Bouchaud}\ and\ \citenamefont
  {Georges}(1990)}]{Bouchaud:PR_195:1990}%
  \BibitemOpen
  \bibfield  {author} {\bibinfo {author} {\bibfnamefont {J.-P.}\ \bibnamefont
  {Bouchaud}}\ and\ \bibinfo {author} {\bibfnamefont {A.}~\bibnamefont
  {Georges}},\ }\href {\doibase 10.1016/0370-1573(90)90099-N} {\bibfield
  {journal} {\bibinfo  {journal} {Physics Reports}\ }\textbf {\bibinfo {volume}
  {195}},\ \bibinfo {pages} {127 } (\bibinfo {year} {1990})}\BibitemShut
  {NoStop}%
\bibitem [{\citenamefont {Frenkel}(1987)}]{Frenkel:PLA_121:1987}%
  \BibitemOpen
  \bibfield  {author} {\bibinfo {author} {\bibfnamefont {D.}~\bibnamefont
  {Frenkel}},\ }\href {\doibase 10.1016/0375-9601(87)90482-8} {\bibfield
  {journal} {\bibinfo  {journal} {Phys. Lett. A}\ }\textbf {\bibinfo {volume}
  {121}},\ \bibinfo {pages} {385} (\bibinfo {year} {1987})}\BibitemShut
  {NoStop}%
\bibitem [{\citenamefont {Olver}\ \emph {et~al.}(2010)\citenamefont {Olver},
  \citenamefont {Lozier}, \citenamefont {Boisvert},\ and\ \citenamefont
  {Clark}}]{Olver:2010:NHMF}%
  \BibitemOpen
  \bibinfo {editor} {\bibfnamefont {F.~W.~J.}\ \bibnamefont {Olver}}, \bibinfo
  {editor} {\bibfnamefont {D.~W.}\ \bibnamefont {Lozier}}, \bibinfo {editor}
  {\bibfnamefont {R.~F.}\ \bibnamefont {Boisvert}}, \ and\ \bibinfo {editor}
  {\bibfnamefont {C.~W.}\ \bibnamefont {Clark}},\ eds.,\ \href@noop {} {\emph
  {\bibinfo {title} {NIST Handbook of Mathematical Functions}}}\ (\bibinfo
  {publisher} {Cambridge University Press},\ \bibinfo {address} {New York,
  NY},\ \bibinfo {year} {2010})\ \bibinfo {note} {print companion to
  \cite{NIST:DLMF}}\BibitemShut {NoStop}%
\bibitem [{{\relax DLMF}()}]{NIST:DLMF}%
  \BibitemOpen
  {\relax DLMF},\ \href {http://dlmf.nist.gov/} {\enquote {\bibinfo {title}
  {{\relax NIST} digital library of mathematical functions},}\ }\bibinfo
  {howpublished} {http://dlmf.nist.gov/, Release 1.0.15 of 2017-06-01},\
  \bibinfo {note} {online companion to \cite{Olver:2010:NHMF}}\BibitemShut
  {NoStop}%
\bibitem [{\citenamefont {Tuck}(1967)}]{Tuck:MoC_21:1967}%
  \BibitemOpen
  \bibfield  {author} {\bibinfo {author} {\bibfnamefont {E.~O.}\ \bibnamefont
  {Tuck}},\ }\href {\doibase 10.2307/2004168} {\bibfield  {journal} {\bibinfo
  {journal} {Math. Comp.}\ }\textbf {\bibinfo {volume} {21}},\ \bibinfo {pages}
  {239} (\bibinfo {year} {1967})}\BibitemShut {NoStop}%
\bibitem [{\citenamefont {Carlson}(1995)}]{Carlson:NA_10:1995}%
  \BibitemOpen
  \bibfield  {author} {\bibinfo {author} {\bibfnamefont {B.~C.}\ \bibnamefont
  {Carlson}},\ }\href {\doibase 10.1007/BF02198293} {\bibfield  {journal}
  {\bibinfo  {journal} {Numerical Algorithms}\ }\textbf {\bibinfo {volume}
  {10}},\ \bibinfo {pages} {13} (\bibinfo {year} {1995})}\BibitemShut {NoStop}%
\bibitem [{\citenamefont {Johansson}\ \emph {et~al.}(2013)\citenamefont
  {Johansson} \emph {et~al.}}]{mpmath}%
  \BibitemOpen
  \bibfield  {author} {\bibinfo {author} {\bibfnamefont {F.}~\bibnamefont
  {Johansson}} \emph {et~al.},\ }\href@noop {} {\emph {\bibinfo {title}
  {mpmath: a {P}ython library for arbitrary-precision floating-point arithmetic
  (version 0.18)}}} (\bibinfo {year} {2013}),\ \bibinfo {note} {{\tt
  http://mpmath.org/}}\BibitemShut {NoStop}%
\bibitem [{\citenamefont {Tinkham}(2003)}]{Tinkham:Dover:2003}%
  \BibitemOpen
  \bibfield  {author} {\bibinfo {author} {\bibfnamefont {M.}~\bibnamefont
  {Tinkham}},\ }\href@noop {} {\emph {\bibinfo {title} {Group Theory and
  Quantum Mechanics}}}\ (\bibinfo  {publisher} {Dover Publications, INC.},\
  \bibinfo {address} {Mineola, New York},\ \bibinfo {year} {2003})\BibitemShut
  {NoStop}%
\end{thebibliography}
%

\end{document}